\documentclass[structabstract]{aa}  
\usepackage{natbib}
\usepackage{graphicx}
\usepackage{lscape}
\usepackage{longtable}
\usepackage{txfonts}
\usepackage{color}                                
                                           
\begin{document}

   \title{The metallicity signature of evolved stars with planets
   \thanks{
     Based on observations made with the Mercator Telescope, operated on the island of La Palma
     by the Flemish Community; and observations made with  the Nordic Optical Telescope, operated
     on the island of La Palma jointly by Denmark, Finland, Iceland, Norway, and Sweden.}\fnmsep 
   \thanks{
     Tables~\ref{parameters_table},~\ref{evolutionary_table},~\ref{tabla_abundancias},
     and ~\ref{kinematic_table} are only available in the electronic version of the paper or
     at the CDS via anonymous ftp to cdsarc.u-strasbg.fr (130.79.128.5)
     or via http://cdsweb.u-strasbg.fr/cgi-bin/qcat?J/A+A/}
     }    


   \author{J. Maldonado 
          \inst{1}
          \and  E. Villaver
          \inst{1}
          \and  C. Eiroa
          \inst{1}  
          }

        \institute{Universidad Aut\'onoma de Madrid, Dpto. F\'isica Te\'orica, M\'odulo 15,
                   Facultad de Ciencias, Campus de Cantoblanco, E-28049 Madrid, Spain}
             

   \offprints{J. Maldonado \\ \email{jesus.maldonado@uam.es}}
   \date{Received 11 January 2013; accepted 12 March 2013}

 
  \abstract
  {}
   {We aim to test whether the well established correlation between the metallicity of
    the star and the presence of giant planets found for main sequence stars still
    holds for the evolved and generally more massive giant and
    subgiant stars. Although several attempts have been made so far, the results
    are not conclusive since they are based on small or inhomogeneous samples. 
   }
   {We determine in a homogeneous way the metallicity and individual abundances of 
    a large sample of evolved stars, with and without known planetary companions, 
    and discuss their metallicity distribution and trends.
    Our methodology is based on the analysis of high-resolution \'echelle spectra
    ($R \ge 67000$) from 2-3 meter class telescopes.
    }  
   {The metallicity distributions 
     show that giant stars hosting planets are not preferentially
     metal-rich having similar abundance patterns to giant stars without known
     planetary companions. We have found, however, a very strong relation between the
     metallicity distribution and the stellar mass within this
     sample. We show that the less massive giant stars with planets (M$\le$ 1.5
     M$_{\odot}$) are not metal rich, but, the metallicity of
     the sample of massive (M$>$ 1.5 M$_{\odot}$), young (age $<$ 2 Gyr) 
     giant stars with planets is higher than that of a similar sample of stars without planets. 
     Regarding other chemical elements, giant stars with and without planets in
     the mass domain M$\le$ 1.5 M$_{\odot}$  show similar abundance
     patterns. However, planet and non-planet hosts with masses M $>$
     1.5 M$_{\odot}$ show differences in the abundances of some elements, specially Na, Co, and Ni.
     In addition,  we find the sample of subgiant stars with
     planets to be metal rich showing similar metallicities to main-sequence planet
     hosts. 
   }
    {
     The fact that giant planet hosts
     in the mass domain M$_{\star}$ $\le$ 1.5 M$_{\odot}$  do not show
     metal-enrichment is difficult to explain.  Given that these
     stars have similar stellar parameters to subgiants and main-sequence
     planet hosts, 
     the lack of the metal-rich signature in low-mass giants could be
     explained if originated from a pollution scenario in the main
     sequence that gets erased as the star become fully
     convective. However, there is no physical reason why it should
     play a role for giants with masses M$_{\star}$ $\le$ 1.5
     M$_{\odot}$ but is not observed for giants with  M$_{\star} >$ 1.5 M$_{\odot}$.
     }


  \maketitle

\section{Introduction}

 Understanding the origin and evolution of planets and planetary systems
 is one of the major goals of modern Astrophysics. Twenty years after
 the discovery of the first exoplanets \citep{1992Natur.355..145W,1995Natur.378..355M},
 we  are still far from understanding which stellar properties influence (and how)
 planet formation the most. Excluding the well established correlation between
 the stellar
 metallicity and the probability that the star hosts a gas-giant planet
 \citep[e.g.][]{2004A&A...415.1153S,2005ApJ...622.1102F},
 any other claim of a chemical trend in planet hosting stars
 has been so far disputed. For instance, the evidence of a higher depletion of lithium in planet host stars
 has been the subject of an intense discussion 
 \citep[e.g.][and references therein]{2009Natur.462..189I,2010A&A...519A..87B,2010ApJ...724..154G,2010A&A...512L...5S},
 as well as whether stars with planets (specially solar analogs)
 show or not different trends on the abundance-condensation temperature 
 \citep[see][and references therein]{2010A&A...521A..33R,2010ApJ...720.1592G,2013arXiv1301.2109G,
 2011MNRAS.416L..80G,2011ApJ...732...55S}.

 The planet-metallicity correlation itself has revealed to be more complex
 than initially thought, as stars with orbiting low-mass planets
  (M$_{\rm p}\sin i$ $<$ 30 M$_{\oplus}$)
 do not seem to be preferentially metal rich
 \citep[][and references therein]{2010ApJ...720.1290G,2011arXiv1109.2497M,2011A&A...533A.141S}.
 This observational result explained within the framework of core-accretion
 models 
 \citep[e.g.][]{1996Icar..124...62P,2003ApJ...598L..55R,2004A&A...417L..25A,2012A&A...541A..97M},
 assumes that the timescale needed to form an icy/rocky core
 is largely dependent on the metal content of the protostellar cloud.
 In this way, in low-metal environments, the gas has already been
 depleted from the disc by the time the cores are massive enough to
 start a runaway accretion of gas and, therefore,
 only low-mass planets can be formed.
 The metallicity patterns found in stars hosting dusty debris discs
 also agree with the predictions of this scenario of planet formation \citep[see][and references therein]{2012A&A...541A..40M}.


  Observations of    
  solar-type (FGK dwarfs) Main Sequence (MS) planet hosts
  point towards a metal rich nature of the MS stars
  throughout their interiors, and therefore, to a primordial nature of the metallicity
  enhancement \citep[e.g.][]{2004A&A...415.1153S,2005ApJ...622.1102F}. 
  Alternative scenarios in which the metal enhancement results from the
  late-stage accretion of H and He-depleted material onto the convective
  zone of the star \citep{1997MNRAS.285..403G,1997ApJ...491L..51L}
  were rapidly ruled-out.
  With our
  current understanding, and given its primordial nature, the observed
  correlation between the metallicity of the star and the presence of
  planets should also hold for Red Giants and Subgiant stars that,
  having left the MS when they exhaust the hydrogen in the core, have larger radii,
  cooler photospheres, and are convective for the most part. 

  The opportunity of
  testing how well-founded the planet-metallicity relation is with a statistically sound
  sample of evolved stars, has become possible recently provided by the large number of
  planets found by the different successful  surveys. Some examples include the
  Lick K-giant Survey \citep{2002ApJ...576..478F},
  the Okayama Planet Search \citep{2003ApJ...597L.157S},
  the Retired A stars and Their Companions \citep{2007ApJ...665..785J},
  or  the Pennsylvania-Toru\'n Planet Search
  \citep{2007ApJ...669.1354N}.

 The first conclusions regarding the
  metallicity of giant stars hosting planets were based on the analysis
  of small or inhomogeneous samples obtained from the different surveys
  available:
  \citet[][with 4 planet-hosting stars analyzed]{2005PASJ...57..127S},
  \citet[][1 star]{2005ApJ...632L.131S}, \citet[][10 stars]{2007A&A...473..979P}.
  These studies suggested that, unlike their MS counterparts, G and K giants 
  stars with planets do not have a tendency to show metal-enrichment.
  An attempt to expand the sample size, setting stellar metallicities from the literature
  in a common spectroscopic scale, is made by 
  \citet[with a total of 20 planet hosts
  analyzed]{2007A&A...475.1003H} where they find evidence that the giant
  planet metallicity correlation might also hold for giants stars.
  More recently, studies based on the analysis of high-resolution spectra,
  \citet[][10 stars]{2008PASJ...60..781T}, and
  \citet[16 stars,][]{2010ApJ...725..721G}, point again towards a lack of a planet-metallicity
  relation for giant stars. This latter study  also included 15 subgiants with planets
  which are found to have, on average, the same metallicity distribution that a sample of dwarf
  stars with planets.

  Evidence that subgiant stars with planets might follow the planet-metallicity
  correlation were previously reported by
  \cite{2005ApJ...622.1102F} who analyzed 9 subgiant stars with planets from a
  total of 1040 stars observed as part of the California \& Carnegie, and the Anglo-Australian
  planet search projects. The metallicity distribution of the planet-host subgiant stars
  appeared to be consistent with that of MS stars with planets, being more metal-rich than
  their counterparts without detected planets.  
  A recent analysis of the California Planet Survey targets is presented in
  \cite{2010PASP..122..905J}
  who analyzed a sample of 1266  stars 
  including a broad range of stellar masses, from late-K and M stars 
  to subgiants with masses up to 1.9 M$_{\odot}$. The authors found evidence of a 
  planet-metallicity correlation for all stellar masses, even when the sample was restricted to 
  subgiant stars with masses in the range M$_{\star}$ $>$ 1.4 M$_{\odot}$ (including
  36 planet hosts). 
  The occurrence of gas-giant planets was found to be not only dependent on the stellar metallicity,
  but it also scales with the stellar mass \citep[see also][]{2011ApJS..197...26J}.

  Several explanations have been put forward to explain the observed metallicity distribution
  of giant planet hosts. For instance, that planets around
  intermediate-mass stars are formed preferentially
  by instabilities, and thus are not dependent on the metallicity of the primordial disk
  \citep[see discussion in][]{2007A&A...473..979P},
  or by late-stage accretion of depleted material onto the convective zone of
  the star \citep{1997MNRAS.285..403G,1997ApJ...491L..51L}. 
  Moreover, recent simulations of planet population synthesis 
  \citep{2011A&A...526A..63A,2012A&A...541A..97M}
  based on the core-accretion model of planet formation, have shown that
  the stellar mass can play a role in planet formation by scaling
  the mass of the protoplanetary disk. In this scenario, 
  a high-mass protoplanetary disk might compensate (at least up to
  certain point) for a 
  low-metallicity environment, allowing the formation of giant-planets even around
  low-metallicity stars.
  The positive correlation found between the presence of gas-giant planets with both stellar metallicity
  and stellar mass \citep[e.g.][]{2010PASP..122..905J}
  could be then explained by assuming that higher mass stars are likely to
  form with more massive protoplanetary disks.

 We believe
 that the analysis of an homogeneous and large sample of evolved stars hosting planets is needed
 before an explanation to the apparent nature of the metallicity correlation
 for evolved stars is invoked. This is precisely the goal of this paper,
 in which we present an homogeneous analysis of a large sample of evolved stars
 based on high resolution and high S/N ratio \'echelle spectra. 

 The paper is organised as follows:
 Section~\ref{secction_observations} describes the stellar samples analysed in this work,
 the spectroscopic observations, and how stellar parameters and abundances are
 obtained. In order to explore the presence of any possible bias
 that could affect our analysis,
 the samples are compared in terms of age, distance, and kinematics, the parameters
 that most likely might affect the metallicity content of a star.
 Possible non-LTE effects are also discussed.
 The metallicity distributions are presented in
 Section~\ref{seccion_resultados}, together with an exploration of the
 parameters that could explain the results, and  the properties of the planets orbiting around 
 evolved stars. The results are
 discussed at length in Section~\ref{seccion_discussion}.
 Our conclusions follow in Section~\ref{conclusions} .

\section{Observations}
\label{secction_observations} 

\subsection{The stellar sample}
\label{stellar_sample}

  Our sample contains 142 evolved stars from which 70
  are known to  host at least one planetary companion
  according to the available data at the Extrasolar Planets
  Encyclopaedia\footnote{http://exoplanet.eu/}. The selection criteria
  of the sample was very simple, from the list of evolved stars with
  confirmed planetary companions we have kept those stars for which a
  high S/N spectra (at least 100) could be taken with the combination
  of instruments and telescopes used. The control sample was drawn
  from the \cite{2008AJ....135..209M}
  list of Hipparcos giants within 100 pc from the Sun, 
  to cover similar stellar parameters as the
  stars with detected planets.

\begin{figure}
\centering
\includegraphics[angle=270,scale=0.45]{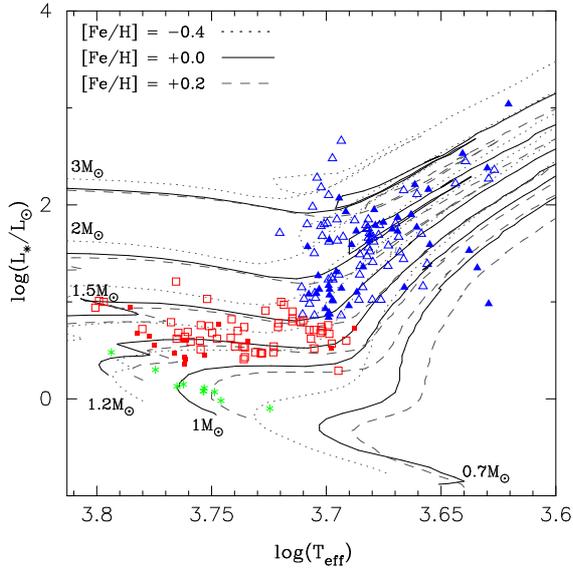}
\caption{
Luminosity versus T$_{\rm eff}$ diagram for the observed stars.
Giants are plotted with blue triangles,
subgiants with red squares, and late main-sequence stars with green asterisks. Filled symbols
indicate planet hosts. Some evolutionary tracks ranging from 0.7 to 3.0 solar masses
from \cite{2000A&AS..141..371G}
are overplotted. For each mass, three tracks are plotted, corresponding to 
Z=0.008 ([Fe/H]=-0.4 dex, dotted lines), Z=0.019 ([Fe/H]=+0.0 dex, solid lines),
and Z=0.030 ([Fe/H]=+0.20 dex, dashed lines).
}
\label{diagrama_hr}
\end{figure}

  Figure~\ref{diagrama_hr}
  shows the HR 
  diagram of the observed stars. They are classified as red giants
  (blue triangles, giants from now on),
  subgiants (red squares), and late MS (green asterisks).
  The classification among the different luminosity classes is somehow
  uncertain for those stars which are in the boundary between two classes.
  In order to distinguish between subgiants and red giants,
  a limit in M$_{\rm bol}$ = 2.82 mag \citep[as in][]{2010ApJ...725..721G}
  was set, although some stars brighter than 2.82 mag which have not yet started their
  ascent into the RGB (Red Giant Branch) have been kept as subgiants.
  In addition, 11 stars which are above the MS tracks on the HR
  diagram, but still have not moved towards the red have been denoted as late MS stars.
  According to their luminosity class and taken into account the presence (or absence)
  of planetary companions, our sample is divided into:
  43 giant stars with known planets (hereafter, GWPs), 67 giant stars without
  planets (GWOPs), 16 subgiants hosting planets (SGWPs), 5 subgiants
  without planets (SGWOPs), and 11 late MS stars harbouring planets
  (LMSWPs). The sample of subgiant stars has been
  supplemented with data from the literature (see Section~\ref{expanding_sgwops}).


\subsection{Spectroscopic observations}
\label{spectroscopic_observations}

 High-resolution spectra of the stars were obtained at La
 Palma observatory (Canary Islands, Spain) during four observing
 runs (two at the MERCATOR telescope and two at the
 Nordic Optical Telescope) between February and August 2011.
 At the MERCATOR telescope  (1.2 m)
 28 stars were observed with the HERMES
 spectrograph \citep{2011A&A...526A..69R}.
 HERMES spectra have a resolution of R $\sim$ 85000 and cover the spectral
 range $\lambda\lambda$ 3800-9000 \AA. HERMES spectra were automatically
 reduced by a detailed data reduction pipeline available at the telescope
 \footnote{See http://www.mercator.iac.es/instruments/hermes/ for details.}.
 The rest of the data, 114 stars, was obtained with the FIES instrument
 \citep{1999anot.conf...71F} at the Nordic Optical Telescope  (2.56 m).
 FIES spectra cover a slightly shorter wavelength range,
 from 3640 to 7360 \space \AA, with a
 resolution of R $\approx$ 67000. FIES spectra were
 reduced using the  {\it advanced} option of the automatic data reduction tool {\it FIEStool}
 \footnote{See http://www.not.iac.es/instruments/fies/fiestool/FIEStool.html for details.}.
 Both pipelines implement the typical corrections involved in \'echelle spectra
 reduction, i.e., bias level, flat-fielding, scattered light correction, removing of
 the blazeshape, order extraction, wavelength calibration, and merge of individual
 orders. HERMES spectra has S/N 
 values between 90 and 340, with an average of $\sim$ 150/160
 in the spectral range around the H$_{\alpha}$ line. 
 In the same spectral range, FIES spectra has a S/N of roughly 75 in the worst
 cases, but up to 480 in the best ones. The average value 
 is around 225.
 The log of the observations is given in Table~\ref{observations_log}.


 The spectra were corrected from radial
 velocity shifts by using the IRAF
 \footnote{IRAF is distributed  by the National Optical Astronomy Observatory, which
 is operated by the Association of Universities for Research in Astronomy, Inc., under
 contract with the National Science Foundation.}
 task {\it dopcor}. Radial velocities were previously measured by cross-correlating
 the spectra of our program stars with spectra of
 radial velocity standard stars of similar spectral types
 obtained during the observations.  


\begin{table}
\centering
\caption{Observing runs performed on 2011.}  
\label{observations_log}
\begin{tabular}{lll}
\hline\noalign{\smallskip}
Date       & Telescope \& Instrument  &  $N$ stars \\
\hline\noalign{\smallskip}
Feb. 14-15 & HERMES/MERCATOR    & 15 \\
May  17-18 & HERMES/MERCATOR    & 13 \\  
May  26-28 & NOT/FIES           & 49 \\
Aug. 16-18 & NOT/FIES           & 65 \\
\hline\noalign{\smallskip}
\end{tabular}
\end{table}

 \subsection{Analysis}
 \label{analysis}

 The basic stellar parameters $T_{\rm eff}$,  $\log g$, microturbulent
 velocity ($\xi_{t}$), and [Fe/H], are determined using the code {\it TGVIT}
 \footnote{http://optik2.mtk.nao.ac.jp/~takeda/tgv/}
 \citep{2005PASJ...57...27T},  which is based on iron-ionization
 and excitation equilibrium  conditions. 

 Iron abundances are computed for a well-defined
 set of 302 {\rm Fe~{I}} and 28 {\rm Fe~{II}} lines. Basically, the stellar
 parameters are adjusted until:  i) no  dependence  is found between the
 abundances derived from  {\rm Fe~{I}} lines and the lower excitation potential
 of the lines; ii) no dependence is found between the abundances derived from
 the  {\rm Fe~{I}} lines and their equivalent widths; and iii) the derived
 mean  {\rm  Fe~{I}}  and  {\rm  Fe~{II}} abundances are the same.
 The line list as well as the adopted parameters (excitation potential,
 $\log$(gf) values, solar EWs) can be found in Y. Takeda's web page. This
 code makes use of ATLAS9, plane-parallel, LTE atmosphere models
 \citep{1993KurCD..13.....K}. The assumed solar Fe abundance
 is A$_{\odot}$ = 7.50 as in \cite{2005PASJ...57...27T}.
 Uncertainties in the stellar parameters
 are computed by progressively changing each stellar
 parameter from the converged solution
 to a value in where any of the aforementioned conditions i), ii), iii) is not longer  
 fulfilled. Uncertainties in the iron abundances are computed by
 propagating the errors in $T_{\rm eff}$,  $\log g$, and $\xi_{t}$.
 We are aware that this procedure only evaluates ``statistical''
 errors. However,  
 other systematic sources of uncertainties such as the choice of the atmosphere
 model, the adopted atomic parameters, or the list lines used are difficult
 to estimate 
 \citep[see for details,][]{2002PASJ...54..451T,2002PASJ...54.1041T}.

 In order to avoid errors due to uncertainties in the damping parameters,
 only lines with EWs $<$ 120 m\AA \space were considered \citep[e.g.][]{2008PASJ...60..781T}.
 Stellar EWs are measured using the automatic code {\it ARES} \citep{2007A&A...469..783S}.
 In order to test the quality of the EWs measured by {\it ARES},
 we selected four representative stars of our sample, covering the whole space
 of parameters, namely
 HIP 118319  (5989 K), 
 HIP 50887   (5001 K), 
 HIP 42527   (4516 K), 
 HIP 100587  (4259 K) 
 and measured the EWs of  iron lines
 ``manually'' by using the  IRAF-task
 {\it splot}.
 Median differences between the measured EWs are:
 $\langle$ EW$_{\rm ARES}$ - EW$_{\rm IRAF}$ $\rangle$ =
 -0.39 $\pm$ 2.1 m\AA, \space
 -0.34 $\pm$ 2.1 m\AA, \space
 -0.48 $\pm$ 2.6 m\AA, \space
 and \space -0.74 $\pm$ 4.1 m\AA \space
 for HIP 118319, HIP 50887, HIP 42527, HIP 118319, respectively.
 We do not find any 
 significant difference between {\it ARES} equivalent widths and the ``manual''
 measurements. The estimated stellar parameters and iron abundances are  given in
 Table~\ref{parameters_table}.


\begin{table*}
\centering
\caption{
 Spectroscopic parameters with uncertainties
 for the stars measured in this work. Columns 7 and 9
 give the mean iron abundance derived from Fe~{\sc I} and Fe~{\sc  II} lines,
 respectively, while columns 8 and 10 give the corresponding number of lines.
 The rest of the columns are self-explanatory. Only the first five lines are
 shown here; the full version of the table is available online.
 }
\label{parameters_table}
\begin{tabular}{llcccccccccl}
\hline\noalign{\smallskip}
HIP &  HD &  T$_{\rm eff}$ &  $\log g$      & $\xi_{t}$     & [Fe/H] &  $\left\langle A({\rm Fe}~{\rm I}) \right\rangle$ & n$_{\rm I}$ &
 $\left\langle A({\rm Fe}~{\rm II}) \right\rangle$  & n$_{\rm II}$ &  Spec.$^{\dag}$ \\
    &     &   (K)   & (${\rm cm s^{-2}}$)  & (${\rm km s^{-1}}$) & dex  &      &      &      &      &       \\
 (1)& (2) &  (3)    & (4)                  & (5)                 & (6)  & (7)  & (8)  & (9)  & (10) & (11)  \\
\hline\noalign{\smallskip}
\multicolumn{11}{c}{Giants with planets}\\
\hline\noalign{\smallskip}
1692	&	1690	&	4343	$\pm$	20	&	2.06	$\pm$	0.08	&	1.56	$\pm$	0.14	&	-0.23	$\pm$	0.04	&	7.27	$\pm$	0.05	&	197	&	7.27	$\pm$	0.07	&	17	&	2	\\
4297	&	5319	&	4900	$\pm$	25	&	3.35	$\pm$	0.09	&	1.10	$\pm$	0.10	&	0.05	$\pm$	0.04	&	7.55	$\pm$	0.04	&	234	&	7.55	$\pm$	0.06	&	18	&	2	\\
10085	&	13189	&	4175	$\pm$	33	&	1.62	$\pm$	0.13	&	1.49	$\pm$	0.17	&	-0.37	$\pm$	0.06	&	7.13	$\pm$	0.07	&	229	&	7.13	$\pm$	0.10	&	20	&	1	\\
12247	&	16400	&	4864	$\pm$	25	&	2.65	$\pm$	0.08	&	1.42	$\pm$	0.10	&	-0.03	$\pm$	0.03	&	7.47	$\pm$	0.04	&	217	&	7.47	$\pm$	0.06	&	17	&	2	\\
	&	17092	&	4634	$\pm$	28	&	2.48	$\pm$	0.10	&	1.31	$\pm$	0.13	&	0.11	$\pm$	0.05	&	7.61	$\pm$	0.05	&	237	&	7.61	$\pm$	0.08	&	21	&	1	\\
\noalign{\smallskip}\hline\noalign{\smallskip}
\multicolumn{11}{l}{$^{\dag}$Spectrograph: (1) MERCATOR/HERMES; (2) NOT/FIES}\\
\end{tabular}
\end{table*}

 \subsection{Photometric parameters and comparison with previous works}
 \label{previous_work}

  Photometric
  effective temperatures are derived from the {\it Hipparcos} 
  $(B-V)$  colours \citep{1997ESASP1200.....P}
  by using the calibration provided by
  \citet[][Table~4]{2010A&A...512A..54C}.
  Uncertainties in the photometric temperatures are estimated by taking
  into account the standard deviation of the calibration ($\sim$ 73 K),
  the uncertainty in the zero point of the temperature scale
  (which is, according to the authors, of the order of 15-20 K), and the 
  propagation of the errors associated with colours and
  metallicities. These three sources of uncertainty have been
  added quadratically. Although this calibration was built using dwarfs
  and subgiants stars, we find that it also reproduces the spectroscopic
  temperatures obtained for our sample of giants.  
  
\begin{figure}
\centering
\includegraphics[angle=270,scale=0.45]{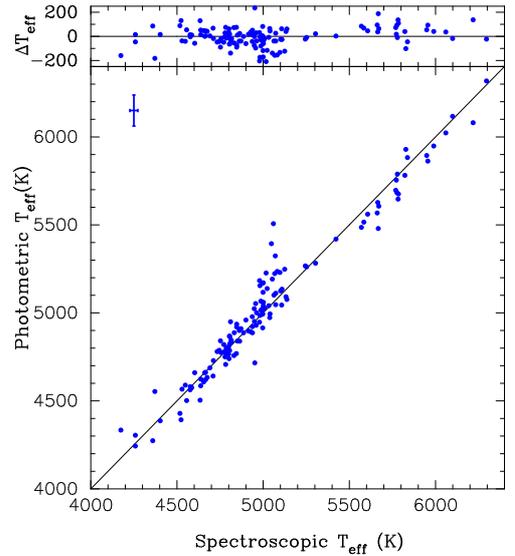}
\caption{ 
Comparison between our spectroscopic-derived T$_{\rm eff}$
and those obtained from $(B-V)$ colours. The upper panel shows the
differences between the spectroscopic and the photometric values.
Mean uncertainties in the derived temperatures are also shown.
}
\label{temperaturas_fotometricas}
\end{figure}

  Since our sample contains stars up to roughly 0.5 kpc,
  colours are de-reddened before we compute the photometric temperatures.
  Visual extinction, A$_{\rm V}$, and colour excesses, E$_{\rm (B-V)}$
  \footnote{The usual relationship A$_{\rm V}$= 3.10 $\times$ E$_{\rm (B-V)}$
   is assumed \citep[e.g.][]{1979ARA&A..17...73S}.},
  are computed as a function of the stellar distance and the galactic
  coordinates  ($l$, $b$) by interpolating in the tables given
  by \cite{1992A&A...258..104A}.
  Distances are obtained from the revised parallaxes provided by 
  \cite{Leeuwen} from a new reduction of the {\it Hipparcos}'s raw data.
  For the five stars with planets that do not have {\it Hipparcos}'s data
  the parallaxes have been taken from the papers in which
  the discovery of the corresponding planets were announced.  
  The comparison between the temperature values obtained by both
  procedures, spectroscopic and photometric, is illustrated in
  Figure~\ref{temperaturas_fotometricas}
  where we do not find any sound systematic difference between
  them,  being the mean value of $\Delta$T${\rm eff}$ $\sim$ -16 K,
  with a standard deviation of only 96 K.  
  We have also computed photometric temperatures using the
  calibration provided by \citet[][Table~5]{2009A&A...497..497G},
  since this relationship was build using giant stars.  
  We note that the temperatures obtained with this relationship
  tend to be slightly cooler than the ones obtained by using 
  the relationship provided by \citet[][]{2010A&A...512A..54C} 
  (and therefore slightly cooler than our spectroscopic values).
  Nevertheless, the difference $\Delta$T${\rm eff}$ is small, 
  $\sim$ 71 K, with a standard deviation of 88 K. 
  The small offset between both calibrations may be  related 
  to the different absolute calibration and zero points
  adopted for Vega \citep{2009A&A...497..497G,2010A&A...512A..54C}.

  Values of the stellar luminosities  ($\log$$L_{\star}/L_{\odot}$)
  are estimated from the absolute magnitudes and
  bolometric corrections using the measurements by \citet[][Table~3]{Flower96}.
  Uncertainties in the stellar luminosities have been computed
  by propagating the errors associated with the $V$ magnitudes,  visual
  extinction, parallaxes, and effective temperatures.
  Estimates of the uncertainty in the visual extinction are
  already given in the tables by \cite{1992A&A...258..104A}, while
  typical uncertainties in $V$ are $\pm$ 0.01 mag
  \citep{1997ESASP1200.....P}. Bolometric corrections have been
  derived as a function of T$_{\rm eff}$. For the error computations,
  the uncertainty due to the propagation
  of the errors in T$_{\rm eff}$, and the sigma of the calibration
  BC-T$_{\rm eff}$ have been added cuadratically. 
  The values of visual extinction, photometric temperatures and
  luminosities are shown in Table~\ref{evolutionary_table}.




\begin{figure}
\centering
\includegraphics[angle=270,scale=0.45]{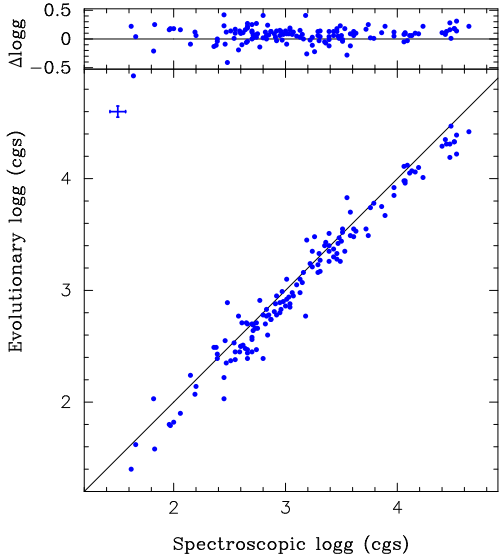}
\caption{ 
Spectroscopic-derived $\log g$ values versus  $\log g$ estimates based on
{\it Hipparcos} parallaxes. The upper panel shows the
differences between the spectroscopic and the {\it Hipparcos} values. 
Mean uncertainties in $\log g$ values are also shown.
}
\label{hipparcos_parallaxes}
\end{figure}

   Evolutionary  
  values of gravities are computed from {\it Hipparcos} $V$ magnitudes
  and parallaxes using L. Girardi's code {\it PARAM}
  \footnote{http://stev.oapd.inaf.it/cgi-bin/param\_1.1}
  \citep{2006A&A...458..609D}, which is based on the use of Bayesian methods.
  Our derived spectroscopic T$_{\rm eff}$ and metallicities are used as inputs
  for {\it PARAM}. The code also estimates the stellar evolutionary parameters,
  age, mass, and radius. These quantities are also given in Table~\ref{evolutionary_table},
  while a comparison between the spectroscopic and evolutionary $\log g$ values is shown
  in Figure~\ref{hipparcos_parallaxes}.
  It is clear from the figure that spectroscopic $\log g$ values tend to be systematically
  larger than the  evolutionary estimates. Specifically, spectroscopic
  values are $\sim$ 0.09 larger (in median) than the evolutionary estimates with
  a standard deviation of 0.13.
  Such a trend of larger spectroscopic $\log g$ values has already been
  reported and discussed by several authors \citep[e.g.][and references therein]{2006A&A...458..609D}
  pointing towards  non-LTE effects on Fe~{\rm I} abundances or thermal inhomogeneities
  as possible causes. We note, however, that the standard deviation
  of the distribution $\log g_{\rm spec}$ - $\log g_{\rm evol}$ is 0.13,
  which is of the same order of magnitude of the uncertainties in the spectroscopic
  derived $\log g$ values. Therefore, we may state that our spectroscopic values
  are in agreement (within the uncertainties) with the  evolutionary estimates,
  ruling out significant departures from LTE conditions
  (see discussion in Section~\ref{possible_biases}).

  There is one outlier, namely BD+20 2457 (left upper corner in
  Figure~\ref{hipparcos_parallaxes}); but this is due to its largely undetermined
  parallax,  $\pi$ = 5.0 $\pm$ 26.0 mas \citep{2009ApJ...707..768N}.


\begin{figure}
\centering
\includegraphics[angle=270,scale=0.45]{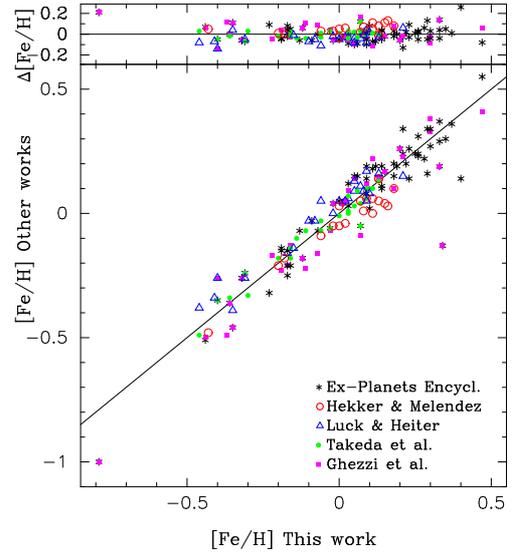}
\caption{
[Fe/H] values, this work, versus literature estimates.
The upper panel shows the differences
between the metallicities derived in this work and the values
given in the literature.}
\label{comparacion_metalicidades}
\end{figure}

 We finally compare our metallicities with those already reported in the literature.
 Values for the comparison are taken from the Extrasolar Planets
 Encyclopaedia\footnote{http://exoplanet.eu/}
 (and references therein) as well as from \cite{2007A&A...475.1003H},
 \cite{2007AJ....133.2464L}, \cite{2008PASJ...60..781T}, and \cite{2010ApJ...725..721G},
 where we were able to find literature metallicities for roughly 70\% of our program stars.
 The comparison is shown in Figure~\ref{comparacion_metalicidades}.
 The agreement is in overall good, with
 $\langle$ [Fe/H]$_{\rm this\, work}$ - [Fe/H]$_{\rm other\, works}$ $\rangle$ = +0.00 dex
 and a standard deviation of 0.08 dex.

\begin{table*}
\centering
\caption{ 
 Photometric and evolutionary parameters for the stars measured
 in this work (see text for details). Only the first five lines are
 shown here; the full version of the table is available online.
 Each quantity is accompanied by its corresponding uncertainty.}
\label{evolutionary_table}
\begin{tabular}{lccccccc}
\hline\noalign{\smallskip}
HIP/       & A$_{\rm V}$ &  $L_{\star}/L_{\odot}$ & T$_{\rm eff}^{\rm phot}$ &  $\log g_{\rm evol}$ &   Age    & Mass           & Radius        \\ 
Other name & (mag)       &  ($\log$)              &        (K)               &  (${\rm cm s^{-2}}$)      &   (Gyr)  & (M$_{\odot}$)  & (R$_{\odot}$) \\ 
 (1)       & (2)         &  (3)                   & (4)                      & (5)                       &   (6)    & (7)            & (8)           \\ 
\hline\noalign{\smallskip}
\multicolumn{8}{c}{Giants with planets}\\
\hline\noalign{\smallskip}
1692	&	0.10	$\pm$	0.04	&	1.53	$\pm$	0.39	&	-			&	1.90	$\pm$	0.10	&	6.72	$\pm$	3.18	&	1.11	$\pm$	0.15	&	18.80	$\pm$	2.77	\\
4297	&	0.10	$\pm$	0.03	&	0.96	$\pm$	0.09	&	4960	$\pm$	86	&	3.40	$\pm$	0.07	&	3.45	$\pm$	0.66	&	1.37	$\pm$	0.08	&	3.72	$\pm$	0.36	\\
10085	&	0.82	$\pm$	0.50	&	3.04	$\pm$	0.41	&	4334	$\pm$	168	&	1.40	$\pm$	0.11	&	4.56	$\pm$	2.97	&	1.19	$\pm$	0.25	&	34.60	$\pm$	6.28	\\
12247	&	0.08	$\pm$	0.06	&	1.73	$\pm$	0.04	&	4839	$\pm$	86	&	2.71	$\pm$	0.05	&	1.38	$\pm$	0.18	&	1.90	$\pm$	0.12	&	9.67	$\pm$	0.40	\\
HD 17092	&	0.20	$\pm$	0.05	&	1.15	$\pm$	0.44	&	4504	$\pm$	86	&	2.89	$\pm$	0.28	&	5.60	$\pm$	3.17	&	1.20	$\pm$	0.20	&	6.34	$\pm$	2.18	\\
\hline\noalign{\smallskip}
\end{tabular}
\end{table*}

 \subsection{Abundance computation}
 \label{abundance_computation}

 Chemical abundances of individual elements (Na, Mg, Al, Si,
 Ca, Sc, Ti~{\sc I}, Ti~{\sc II}, Mn, Cr~{\sc I}, Cr~{\sc II},
 V, Co, Ni, Zn)
 are obtained by using the {\it WIDTH9} program \citep{2005MSAIS...8...44C}
 together with ATLAS9 atmosphere models \citep{1993KurCD..13.....K},
 updated to work under Linux by \cite{2004MSAIS...5...93S} and
 \cite{2005MSAIS...8...61S}.

 The measured equivalent widths of a list of narrow, non-blended lines for each
 of the aforementioned ions are used as inputs for {\it WIDTH9}. The selected
 lines are mainly taken from the list provided by \citet[][Table~2]{2009A&A...497..563N}, although
 we keep the parameters of the lines (excitation potential, oscillator strength) as given
 in Kurucz's lists of lines. 
For Zn abundances, the lines at 4810.54 and
 6362.34 \AA \space were considered.

 The abundances obtained are given in Table~\ref{tabla_abundancias}. 
 They are expressed relative to the
 solar values provided by \cite{2009ARA&A..47..481A}. 
 We have used the
 four representative stars mentioned in Section~\ref{analysis} in 
 order to provide an estimate on how the uncertainties in the
 atmospheric parameters propagate into the abundance calculation. 
 Abundances for each of these four stars have been recomputed using
 T$_{\rm eff}$ + $\Delta$T$_{\rm eff}$, T$_{\rm eff}$ - $\Delta$T$_{\rm eff}$,
 and similarly for $\log g$ and $\xi_{t}$. Results are given in 
 Table~\ref{abundance_sensitivity}. 
 As final uncertainties for the derived abundances we give the quadratically
 sum of the uncertainties due to the propagation of the errors in the stellar
 parameters, plus 
 the line-to-line scatter errors
 (computed as $\sigma/\sqrt{\rm N}$, where $\sigma$ is the standard deviation of the derived individual
 abundances from the $N$ lines). 
 We would like to point out here that even these uncertainties should be considered as 
 lower limits, given that the errors in the stellar parameters are only
 statistical (as explained in Section~\ref{analysis}), and the abundance estimates are affected by
 systematics which are not
 taken into account in line-to-line errors 
 (i.e.  atomic data, or uncertainties in the atmosphere models).

\begin{figure*}
\centering
\includegraphics[angle=270,scale=0.75]{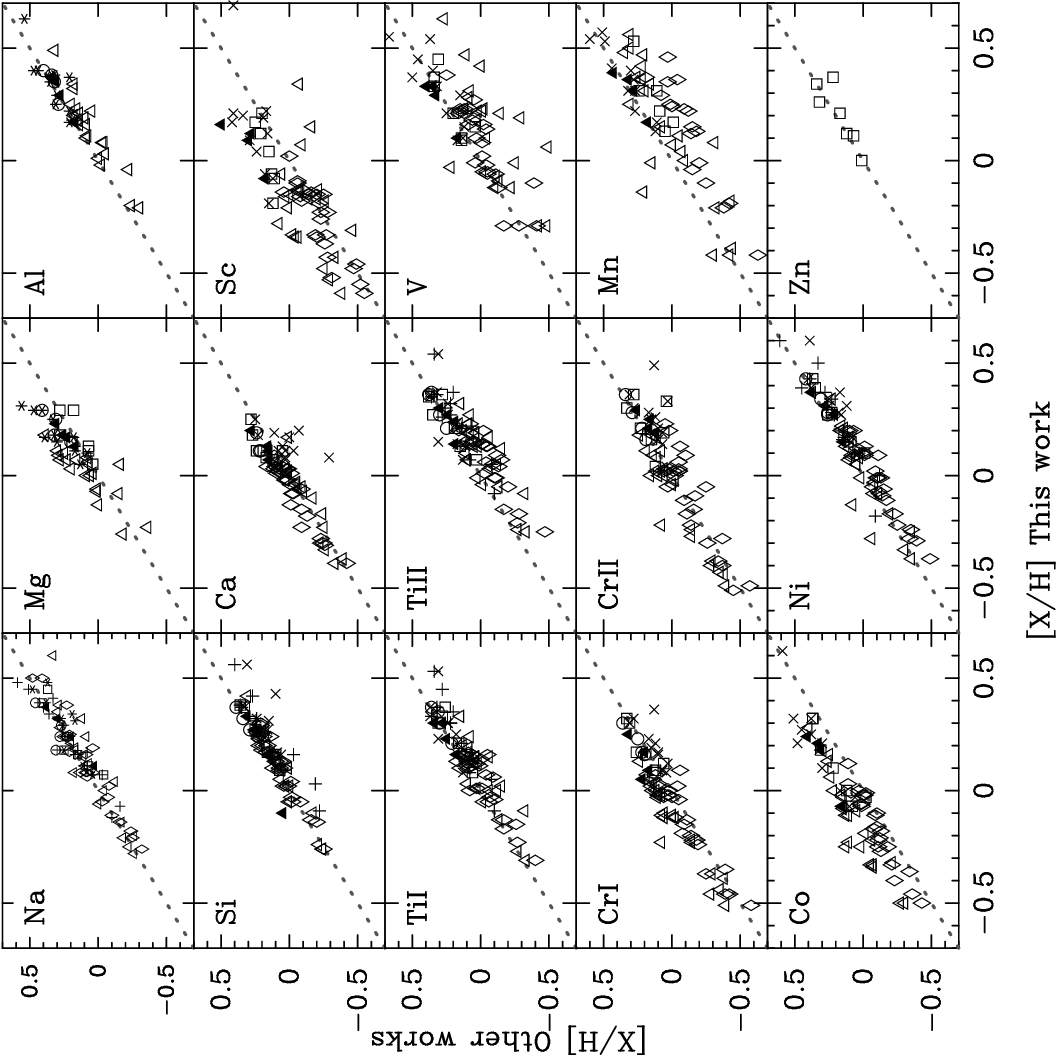}
\caption{
Comparison of our abundances to those of
\cite{2005A&A...438..251B}($\ast$), \cite{2005A&A...433..185B} (open circles), \cite{2005ApJS..159..141V}($+$),
\cite{2006A&A...449..723G}($\times$), \cite{2007AJ....133.2464L} (open triangles), \cite{2007PASJ...59..335T} (open squares)
\cite{2008PASJ...60..781T} (diamonds), and \cite{2009A&A...497..563N} (filled triangles).}
\label{previous_abund}
\end{figure*}

 A comparison of our derived abundances with those previously reported in
 the literature is shown in Figure~\ref{previous_abund}.
 Derived abundances of Na, Al, Ti, and Ni 
 agree reasonable well with previously reported values, with the
 $\sigma$ of the distribution [X/H](this work) - [X/H](other work) ranging from $\sim$ 0.03
 to 0.08 dex, although our abundances seem to be slightly shifted towards 
 higher values (maximum mean differences $\le$ $\sim$ 0.08 dex).  
 In the cases of Si and Ca, our abundances are in median a bit larger ($\sim$ 0.1 dex) 
 than those given in \cite{2006A&A...449..723G} and
 \cite{2005ApJS..159..141V}, but in agreement with 
 \cite{2007AJ....133.2464L,2007PASJ...59..335T,2008PASJ...60..781T}.
 For Mg, our abundances are on average a bit lower (within 0.1 dex) than those given
 by \cite{2005ApJS..159..141V} and \cite{2006A&A...449..723G}, although
 in excellent agreement with \cite{2007PASJ...59..335T}.
 Abundances of Cr, and Co are slightly lower than those previously reported,
 specially in the case of Co, but still mean differences are within $\pm$ 0.1 dex.
 The largest dispersion are found for Sc, V, and Mn, probably due to the
 small number of lines used for these elements or uncertainties in the 
 atomic parameters.
  It is well known that some lines of Sc, V, Mn (and also Co) 
  split into different subcomponents due to electron-nucleus interactions
  showing a significant hyperfine structure, \citep[e.g.][]{2011ApJ...732...55S}.
  Hyperfine structure (hfs) has not been considered in our analysis
  and, as a consequence the abundances of these elements may be overestimated. 
  We note, however, that the  differences between hfs synthesis abundances and
  EW-based abundances derived by \cite{2011ApJ...732...55S} are small,
  $\leq$ 0.04 dex, in 8 out of the 10 late-F and G type analysed stars.
  In addition, we do not
  expect hfs effects to bias the results of the comparisons performed
  in this work (see Section~\ref{discussion_pollution2})  between
  samples of stars with and without planets, given that they have otherwise similar properties. 
  
  Finally, considering Zn, only \cite{2008PASJ...60..781T} give abundances for this element. Despite the small number
  of stars in common the agreement is quite clear, as shown in Figure~\ref{previous_abund}.

\begin{table}
\centering
\caption{
Abundance sensitivities.
}
\label{abundance_sensitivity}
\begin{scriptsize}
\begin{tabular}{lrrrrrr}
\hline\noalign{\smallskip}
               &  \multicolumn{3}{c}{HIP 118319}    & \multicolumn{3}{c}{{HIP 50887} } \\
Ion            &  \multicolumn{3}{c}{\hrulefill}      				  & \multicolumn{3}{c}{\hrulefill}         \\
               & $\Delta$T$_{\rm eff}$ &  $\Delta$$\log g$  &  $\Delta$$\xi_{t}$  & $\Delta$T$_{\rm eff}$  & $\Delta$$\log g$   & $\Delta$$\xi_{t}$   \\
               & $\pm$25               &  $\pm$0.05         &  $\pm$0.14          & $\pm$10                & $\pm$0.04          & $\pm$0.06           \\   
               &  (K)                  &  (cms$^{\rm -2}$)  &  (kms$^{\rm -1}$)   & (K)                    &  (cms$^{\rm -2}$)  &  (kms$^{\rm -1}$)   \\
\hline\noalign{\smallskip}
Na             & 0.01 &   0.01 &     0.05 &  $<$0.01 &   $<$0.01  &    0.01 \\
Mg             & 0.01 &   0.01 &     0.13 &  $<$0.01 &   $<$0.01  &    0.01 \\
Al             & 0.01 &   $<$0.01 &     0.03 &  0.01 &   $<$0.01  &    0.01 \\
Si             & 0.01 &   $<$0.01 &     0.02 &  $<$0.01 &   $<$0.01  &    $<$0.01 \\
Ca             & 0.01 &   0.01 &     0.16 &  0.01 &   $<$0.01  &    0.03 \\
Sc             & 0.03 &   $<$0.01 &     0.05 &  0.01 &   $<$0.01  &    0.01 \\
T~{\sc I}      & 0.02 &   $<$0.01 &     0.03 &  0.02 &   $<$0.01  &    0.02 \\
Ti~{\sc II}    & $<$0.01 &   0.02 &     0.05 &  $<$0.01 &   0.02  &    0.02 \\
V              & 0.02 &   $<$0.01 &     0.15 &  0.01 &   $<$0.01  &    0.02 \\
Cr~{\sc I}     & 0.03 &   0.01 &     0.11 &  0.01 &   $<$0.01  &    0.02 \\
Cr~{\sc II}    & $<$0.01 &   0.02 &     0.20 &  0.01 &   0.02  &    0.02 \\
Mn             & 0.02 &   $<$0.01 &     0.21 &  0.01 &   $<$0.01  &    0.03 \\
Co             & 0.02 &   $<$0.01 &     0.05 &  0.01 &   0.01  &    0.02 \\
Ni             & 0.01 &   $<$0.01 &     0.08 &  0.01 &   0.01  &    0.01 \\
Zn             & 0.01 &   $<$0.01 &     0.28 &  $<$0.01 &   0.01  &    0.02 \\
\hline\noalign{\smallskip}
               &  \multicolumn{3}{c}{HIP 42527}                          & \multicolumn{3}{c}{HIP 100587} \\
Ion            &  \multicolumn{3}{c}{\hrulefill}                                  & \multicolumn{3}{c}{\hrulefill}         \\
               & $\Delta$T$_{\rm eff}$ & $\Delta$$\log g$   & $\Delta$$\xi_{t}$  & $\Delta$T$_{\rm eff}$  & $\Delta$$\log g$  & $\Delta$$\xi_{t}$  \\
               & $\pm$18               & $\pm$0.07          &  $\pm$0.09         & $\pm$35                & $\pm$0.13         &  $\pm$0.15         \\
               &  (K)                  &  (cms$^{\rm -2}$)  &  (kms$^{\rm -1}$)  & (K)                    & (cms$^{\rm -2}$)  &  (kms$^{\rm -1}$)  \\
\hline\noalign{\smallskip}
Na             & 0.01  &  0.01 &     0.03  &  0.03 &   0.02 & 0.06  \\
Mg             & 0.01  &  0.01 &     0.02  &  0.01 &   0.01 & 0.04  \\
Al             & 0.01  &  $<$0.01 &     0.02  &  0.02 &   $<$0.01 & 0.04  \\
Si             & 0.01  &  0.01 &     0.01  &  0.04 &   0.04 & 0.03  \\
Ca             & 0.02  &  0.01 &     0.05  &  0.03 &   0.01 & 0.09  \\
Sc             & 0.02  &  $<$0.01 &     0.03  &  0.04 &   $<$0.01 & 0.11  \\
T~{\sc I}      & 0.03  &  $<$0.01 &     0.06  &  0.05 &   0.01 & 0.12  \\
Ti~{\sc II}    & $<$0.01  &  0.03 &     0.05  &  0.01 &   0.06 & 0.08  \\
V              & 0.03  &  0.01 &     0.08  &  0.04 &   0.01 & 0.17  \\
Cr~{\sc I}     & 0.02  &  $<$0.01 &     0.04  &  0.03 &   $<$0.01 & 0.08  \\
Cr~{\sc II}    & 0.02  &  0.03 &     0.03  &  0.03 &   0.06 & 0.05  \\
Mn             & 0.01  &  $<$0.01 &     0.07  &  0.01 &   $<$0.01 & 0.13  \\
Co             & 0.01  &  0.01 &     0.04  &  $<$0.01 &   0.04 & 0.08  \\
Ni             & $<$0.01  &  0.02 &     0.03  &  0.01 &   0.04 & 0.07  \\
Zn             & 0.02  &  0.02 &     0.04  &  0.03 &   0.03 & 0.04  \\
\hline\noalign{\smallskip}
\end{tabular}    
\end{scriptsize} 
\end{table}   
   

\begin{table*}
\centering
\caption{
Derived abundances of Na, Mg, Al, Si, Ca, Sc, Ti~{\sc i}, Ti~{\sc ii}, V, Cr~{\sc i}, Cr~{\sc ii}, Mn, Co, Ni, and Zn.
Only the first five lines are shown here; the full version of the table is available online.
}
\label{tabla_abundancias}
\begin{scriptsize}
\begin{tabular}{lrrrrrrrrrrrrrrr}
\hline\noalign{\smallskip}
HIP/Other & [Na/H]& [Mg/H] &  [Al/H] & [Si/H] & [Ca/H] & [Sc/H] & [Ti~{\sc i}/H] & [Ti~{\sc ii}/H] & [V/H] &  [Cr~{\sc i}/H] &  [Cr~{\sc ii}/H] & [Mn/H] & [Co/H] & [Ni/H] & [Zn/H] \\
\hline\noalign{\smallskip}
\multicolumn{16}{c}{Giants with planets}\\
\hline\noalign{\smallskip}
1692         &       -0.12 &       -0.17 &        0.10 &        0.07 &       -0.34 &       -0.11 &        0.12 &        0.12 &        0.23 &       -0.30 &       -0.20 &        0.05 &       -0.20 &       -0.12 &       -0.22\\
 & $\pm$        0.11 & $\pm$        0.06 & $\pm$        0.04 & $\pm$        0.12 & $\pm$        0.12 & $\pm$        0.27 & $\pm$        0.15 & $\pm$        0.22 & $\pm$        0.24 & $\pm$        0.10 & $\pm$        0.13 & $\pm$        0.18 & $\pm$        0.17 & $\pm$        0.09 & $\pm$        0.29\\
4297         &        0.13 &        0.05 &        0.24 &        0.15 &       -0.01 &       -0.05 &        0.15 &        0.09 &        0.32 &       -0.02 &        0.01 &        0.33 &        0.02 &        0.12 &        0.12\\
 & $\pm$        0.06 & $\pm$        0.06 & $\pm$        0.03 & $\pm$        0.06 & $\pm$        0.10 & $\pm$        0.13 & $\pm$        0.04 & $\pm$        0.09 & $\pm$        0.15 & $\pm$        0.04 & $\pm$        0.08 & $\pm$        0.14 & $\pm$        0.11 & $\pm$        0.04 & $\pm$        0.20\\
10085        &       -0.11 &       -0.21 &       -0.07 &       -0.01 &       -0.36 &       -0.09 &        0.15 &       -0.17 &        0.35 &       -0.20 &       -0.40 &       -0.05 &       -0.29 &       -0.19 &       -0.56\\
 & $\pm$        0.09 & $\pm$        0.10 & $\pm$        0.12 & $\pm$        0.09 & $\pm$        0.12 & $\pm$        0.27 & $\pm$        0.16 & $\pm$        0.16 & $\pm$        0.24 & $\pm$        0.12 & $\pm$        0.14 & $\pm$        0.23 & $\pm$        0.16 & $\pm$        0.09 & $\pm$        0.21\\
12247        &        0.16 &        0.05 &        0.11 &        0.13 &       -0.04 &       -0.23 &        0.06 &        0.04 &        0.08 &       -0.08 &       -0.15 &        0.22 &       -0.16 &       -0.01 &        0.02\\
 & $\pm$        0.06 & $\pm$        0.11 & $\pm$        0.03 & $\pm$        0.06 & $\pm$        0.10 & $\pm$        0.12 & $\pm$        0.04 & $\pm$        0.10 & $\pm$        0.10 & $\pm$        0.04 & $\pm$        0.06 & $\pm$        0.12 & $\pm$        0.11 & $\pm$        0.04 & $\pm$        0.06\\
HD 17092      &        0.42 &        0.11 &        0.33 &        0.34 &        0.08 &        0.05 &        0.14 &        0.15 &        0.43 &        0.05 &        0.05 &        0.46 &        0.11 &        0.25 &        0.69\\
 & $\pm$        0.22 & $\pm$        0.06 & $\pm$        0.04 & $\pm$        0.08 & $\pm$        0.10 & $\pm$        0.20 & $\pm$        0.09 & $\pm$        0.19 & $\pm$        0.18 & $\pm$        0.06 & $\pm$        0.09 & $\pm$        0.22 & $\pm$        0.14 & $\pm$        0.06 & $\pm$        0.13\\
\hline\noalign{\smallskip}
\end{tabular}
\end{scriptsize}
\end{table*}

\subsection{Expanding the SGWOP sample}
\label{expanding_sgwops}

 Given the small number of stars observed classified as SGWOPs, we
 have expanded the sample with data from the literature in order to make
 a proper comparison between the properties of subgiants with and
 without planetary companions. 
 We have added to the SGWOP sample those stars given in \citet[][hereafter VF05]{2005ApJS..159..141V}
 which  fulfilled our criteria for being classified as subgiants
 (Section~\ref{stellar_sample}).
 These stars have been monitored for planets on the Keck, Lick, and
 Anglo-Australian Telescope planet search programs, discarding the presence
 of planetary companions with radial velocity semiamplitudes $K$ $>$ 30 m s$^{\rm -1}$
 and orbital periods shorter than 4 yr \citep{2005ApJ...622.1102F}.   
 Those stars already observed by us, as well as stars with recently
 discovered planets, were discarded. The final number of stars added to the SGWOP sample
 amounts to 50.


 In order to keep the analysis as homogeneous as possible, VF05 metallicities were
 set into our own metallicity scale by using the stars in common. A linear fit
 was made, obtaining the following linear transformation:
 [Fe/H](our scale) = (0.96$\pm$0.11)$\times$[Fe/H](VF05) - (0.04$\pm$0.03),
 (RMS=0.07, $\chi^{2}_{r}$ $\sim$ 10.4).
 Effective temperatures provided by VF05 were also set into our own
 temperature scale by using the linear relationship:
 T$_{\rm eff}$(our scale) = (1.02$\pm$0.03)$\times$T$_{\rm eff}$(VF05) - (140$\pm$179),
 (RMS=42, $\chi^{2}_{r}$ $\sim$ 5.25).
 Considering $\log g$ values, we get the following transformation:
 $\log g$(our scale) = (1.17$\pm$0.06)$\times$$\log g$(VF05) - (0.79$\pm$0.26),
 (RMS=0.009, $\chi^{2}_{r}$ $\sim$ 2.2).
 
 Stellar ages, masses and radius for these stars were recomputed following
 the same procedure that the one used for the stars analyzed in this work
 (Section~\ref{previous_work}), 
 and are also listed in Table~\ref{evolutionary_table}. 

\subsection{Possible biases}
\label{possible_biases}

 Before we proceed further in the comparison between the different
 samples it is due an exploration of the possible sources of bias that
 could mimic metallicity differences.  Metallicity reflects the enrichment history of the
 ISM \citep[see e.g.][]{1995ApJS...98..617T}. 
 It is, therefore, important
 to determine whether the different 
 samples have randomly selected stellar hosts in terms of age, distance, and
 kinematics, which are the parameters most likely to reflect the
 original metal
 content of the molecular cloud where the stars were born.
 The properties of the stars
 obtained with the procedure explained in the previous subsections,
 are summarised for the different samples in Table~\ref{bias_table}.

\begin{figure}
\centering
\includegraphics[angle=270,scale=0.45]{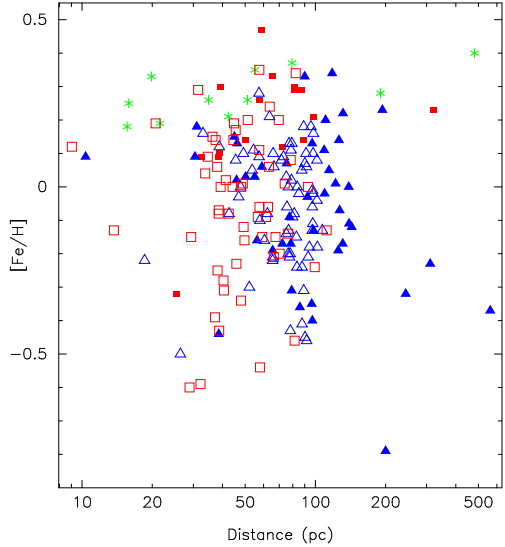}
\caption{[Fe/H] as a function of the stellar distance. Colours and symbols are as in
Figure~\ref{diagrama_hr}. }
\label{metalicidad_distancia}
\end{figure}

 The comparison of the stellar properties among the different samples
 show that planet hosts tend to be systematically at larger
 distances than the stars without known planetary companions. This is
 no unexpected since we selected the control sample from stars within 100 pc,
 while the sample of stars with planets is not volume limited. In order to check whether there
 is a systematic trend in the metallicity due to the distance, the [Fe/H]-distance space is shown
 in Figure~\ref{metalicidad_distancia}. 
 It can be seen  from the figure that GWPs and GWOPs located within $\sim$ 100 pc are well-mixed
 in the [Fe/H]-distance plane showing a similar behaviour. At slightly
 larger distances and up to 200 pc the GWP sample covers approximately the
 same range in [Fe/H] as the sample within  $\sim$ 100 pc. However,
 the four GWPs located beyond 200 pc have very small negative metallicities, specially BD+20 2457  (already mentioned in
 Section~\ref{previous_work}). 
 We consider this figure (four stars, $\sim$ 9\% of the whole GWP sample) too-low
 to bias the metallicity distribution of the GWP sample, so we do not expect any
 significant chemical difference between the GWP and GWOP samples introduced by
 their distances from the Sun.  Nevertheless, we have checked whether
   this is indeed the case in Section~\ref{mass_dependence}.


\begin{table}
\centering
\caption{
Comparison between the properties of the different samples studied in this work.
}
\label{bias_table}
\begin{scriptsize}
\begin{tabular}{lllllll}
\hline\noalign{\smallskip}
               &  \multicolumn{3}{c}{\textbf{GWOPs}}  & \multicolumn{3}{c}{\textbf{GWPs }} \\
               &  \multicolumn{3}{c}{\hrulefill}      & \multicolumn{3}{c}{\hrulefill}      \\
               & Range       & Mean        &  Median  &   Range       & Mean     & Median    \\
\hline\noalign{\smallskip}
 V (mag)           & 2.8/7.8     & 5.5    &  5.5     & 1.1/9.8       &   6.2    &  6.1  \\
 Distance (pc)     & 18.5/107.2  & 75.4   & 78.1     & 10.4/561.8    & 112.3    & 96.9  \\
 Age (Gyr)         & 0.2/9.8     & 2.6    &  1.9     & 0.4/10.5      &   3.0    &  2.4  \\
 T$_{\rm eff}$ (K) & 4235/5252   & 4850   & 4847     & 4175/5107     &  4779    & 4861  \\
 M (M$_{\odot}$)   & 0.9/3.8     & 1.8    & 1.6      & 0.9/2.9       &   1.6    &  1.5  \\
\hline\noalign{\smallskip}
 SpType (\%)         & \multicolumn{3}{l}{ 36 (G); 64 (K)}         & \multicolumn{3}{l}{33 (G); 67 (K)} \\
 D/TD$^{\dag}$ (\%)  & \multicolumn{3}{l}{ 84 (D); 1 (TD); 15 (R)} & \multicolumn{3}{l}{79 (D); 5 (TD); 16 (R)} \\
\noalign{\smallskip}\hline\noalign{\smallskip}
               &  \multicolumn{3}{c}{\textbf{SGWOPs}}  &  \multicolumn{3}{c}{\textbf{SGWPs}}   \\
               &  \multicolumn{3}{c}{\hrulefill}       &  \multicolumn{3}{c}{\hrulefill}       \\
               & Range       & Mean        &  Median   &   Range       & Mean     & Median     \\
\hline\noalign{\smallskip}
 V (mag)           & 3.5/8.6    &  6.6  & 6.6   & 4.5/10.5     & 7.4    &  8.0     \\
 Distance (pc)     & 9.0/112    & 51.4  & 50.0  & 25.3/320.5   & 77.3   & 65.5     \\
 Age (Gyr)         & 1.9/11.7   &  5.3  & 4.3   & 0.9/7.6      & 4.8    &  4.9     \\
 T$_{\rm eff}$ (K) & 4913/6318  & 5431  & 5382  & 4873/6566    & 5745   & 5779     \\
 M (M$_{\odot}$)   & 1.0/1.6    & 1.2   & 1.2   & 1.1/1.5     & 1.2    & 1.2      \\
\hline\noalign{\smallskip}
 SpType (\%)       & \multicolumn{3}{l}{5.5 (F); 74.5 (G); 20 (K)} & \multicolumn{3}{l}{12.5 (F); 75 (G); 12.5 (K)}  \\
 D/TD$^{\dag}$ (\%)& \multicolumn{3}{l}{62 (D); 5 (TD); 33 (R)}    & \multicolumn{3}{l}{56 (D); 6 (TD); 38 (R)}  \\
\noalign{\smallskip}\hline\noalign{\smallskip}
                   &  \multicolumn{3}{c}{\textbf{LMSWPs }} & \multicolumn{3}{c}{}  \\
                   &  \multicolumn{3}{c}{\hrulefill}       & \multicolumn{3}{c}{}  \\
                   &  Range       & Mean     & Median      & \multicolumn{3}{c}{}  \\
\hline\noalign{\smallskip}
 V (mag)           &  5.5/12.2      &   8.0    &  7.9  & \multicolumn{3}{c}{}  \\
 Distance (pc)     &  15.6/480.8    &  91.5    & 42.3  & \multicolumn{3}{c}{}  \\
 Age (Gyr)         &  0.5/10.6      &   4.7    &  4.7  & \multicolumn{3}{c}{}  \\
 T$_{\rm eff}$ (K) &  5304/6597     &  5805    & 5671  & \multicolumn{3}{c}{}  \\
 M (M$_{\odot}$)   &  0.9/1.4       &   1.1    &  1.0  & \multicolumn{3}{c}{}  \\
\hline\noalign{\smallskip}
 SpType (\%)       &  \multicolumn{3}{l}{9 (F);  82 (G); 9 (K)} & \multicolumn{3}{c}{}  \\
 D/TD$^{\dag}$ (\%)&  \multicolumn{3}{l}{82 (D); 18 (R)}         & \multicolumn{3}{c}{}  \\
\noalign{\smallskip}\hline\noalign{\smallskip}
\multicolumn{7}{l}{$^{\dag}$ D: Thin disc, TD: Thick disc, R: Transition} \\
\end{tabular}
\end{scriptsize}
\end{table}


 Thick disc stars are expected to be relatively old
 \citep[e.g][]{2005A&A...433..185B}, metal-poor, and to show $\alpha$-enhancement
 \citep[e.g][]{1998A&A...338..161F,2008MNRAS.388.1175H,2011A&A...535L..11A}. 
 We have
 checked  if there are differences between the different samples
 in terms of membership to the thin/thick disc. The procedure involves
 measuring radial velocities by
 cross-correlating the spectra of the stars with spectra of radial velocity standard
 stars of similar spectral types.  For the SGWOPs stars taken from the literature,
 the radial velocities values  have been mainly taken from the compilation by \cite{2007AN....328..889K}.
 
 Galactic spatial-velocity components ($U, V, W$)
 are computed from the radial velocities, together with {\it Hipparcos} parallaxes
 \citep{Leeuwen}, and {\it Tycho-2} proper motions \citep{2000A&A...355L..27H}, following
 the procedure described in \cite{2001MNRAS.328...45M}, and 
 \cite{2010A&A...521A..12M}. 

\begin{figure}
\centering
\includegraphics[angle=270,scale=0.45]{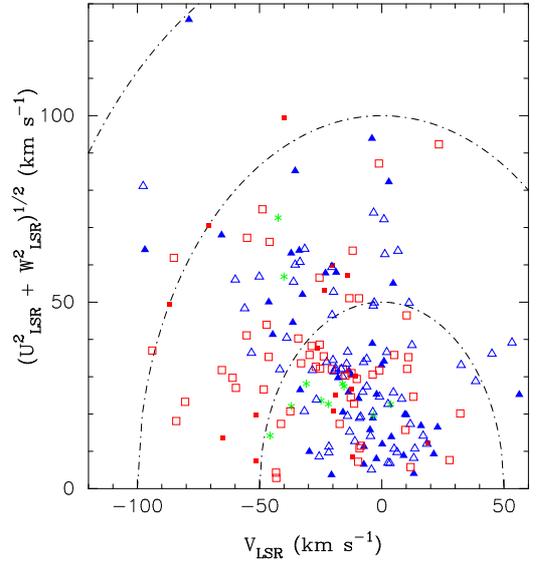}
\caption{
Toomre diagram of the observed stars. Colours and symbols are as in
Figure~\ref{diagrama_hr}. Dash-dot lines indicate constant total velocities,
$V_{\rm Total} = \sqrt{U^{2}_{\rm LSR} +V^{2}_{\rm LSR} + W^{2}_{\rm LSR}}$ = 50, 100, and 150 kms$^{\rm -1}$.
}
\label{toomre_diagrama}
\end{figure}

 For stars in known binary systems the radial velocity of the centre of mass
 of the system is used.  
 Finally, stars have been classified as belonging to the
 thin/thick disc applying the methodology described in
 \cite{2003A&A...410..527B,2005A&A...433..185B}.
 Figure~\ref{toomre_diagrama} 
 shows the Toomre diagram for the observed stars, while the derived velocities are
 given in Table~\ref{kinematic_table}. This type of diagram constitutes a useful way to 
 discriminate stellar populations in velocity space, since 
 it plots the energy versus the the angular
 momentum properties of the stars \citep[e.g.][]{2004AN....325....3F}. 
 We find that roughly $\sim$
 80\% of the stars belong to the thin disc and we do not find any difference in the distribution of the different samples;
 in particular, there are no differences between planet host and stars
 without planets.  It is worth mention that while our classification
 of thin/thick disc stars is based only on kinematical
 criteria, a complete description of the thin/thick disc populations would
 require the combination of kinematics, metallicities, and stellar ages
 \citep[e.g.][]{1998A&A...338..161F}. Nevertheless, the methodology
 used is sufficient to discard the presence of a significant fraction of thick disc stars within any of our samples.

 We note that the two GWP stars possible members of the thick disk have low metallicities
 ([Fe/H] $<$ -0.3 dex)   
and $\sim$ 43\% of the GWPs classified as transition stars
 have also metallicities below -0.3 dex. \cite{2008A&A...482..673H} argued that at metallicities [Fe/H] $<$ -0.3 dex,
 giant planets seem to favour thick disk stars. Statistics of thick
 disc stars are too small in our sample, but they do not contradict \cite{2008A&A...482..673H} idea.


\begin{table}[!htb]
\centering
\caption{Radial velocities and Galactic spatial-velocity components
 for the observed stars. Only the first five lines are
 shown here; the full version of the table is available online.
 The assumed solar motion with respect to the LSR is
 ($U_{\odot}$, $V_{\odot}$, $W_{\odot}$) = (10.0, 5.25, 7.17) km s$^{\rm -1}$ \citep{1998MNRAS.298..387D}.}
\label{kinematic_table}
\begin{scriptsize}
\begin{tabular}{lrrrrc}
\hline\noalign{\smallskip}
HIP/          & V$_{\rm r}$$^{\star}$   &  $U_{\rm LSR}$         & $V_{\rm LSR}$            & $W_{\rm LSR}$        &  C$^{\dag}$ \\  
Other         & (kms$^{-1}$)	 	&  (kms$^{-1}$)          & (kms$^{-1}$)             & (kms$^{-1}$)         &             \\
 (1)          & (2)          		&  (3)                   & (4)                      & (5)                  &   (6)       \\
\hline\noalign{\smallskip}
\multicolumn{6}{c}{Giants with planets}\\
\hline\noalign{\smallskip}
1692	&	17.58	$\pm$	0.25	&	-8.80	$\pm$	8.15	&	  3.94	$\pm$	3.75	&      -10.75	$\pm$	0.94	&	D	\\
4297	&	-0.21	$\pm$	0.31	&	25.28	$\pm$	1.68	&	-13.56	$\pm$	2.00	&	-5.40	$\pm$	1.33	&	D	\\
10085	&	25.36	$\pm$	0.46	&      -13.81	$\pm$	3.55	&	 22.85	$\pm$	3.10	&	 8.97	$\pm$	6.33	&	D	\\
12247	&	8.60	$\pm$	0.31	&        3.65	$\pm$	0.49	&	-20.57	$\pm$	1.33	&	-0.63	$\pm$	0.41	&	D	\\
HD17092 &       5.56    $\pm$   0.31    &       -7.50   $\pm$   6.73    &       -7.26   $\pm$   8.17    &       8.42    $\pm$   1.49    &       D       \\
\hline\noalign{\smallskip}
\multicolumn{6}{l}{$^{\star}$ For those stars in binary systems we have considered the radial velocity of the centre of} \\
\multicolumn{6}{l}{mass of the system.} \\
\multicolumn{6}{l}{$^{\dag}$ Thin/thick disc classification, D: Thin disc, TD: Thick disc, R: Transition} \\
\end{tabular}
\end{scriptsize}
\end{table}

 Both samples of giant stars, with and without planets, cover a wider
 stellar mass range and represent on average a younger population of
 stars than the subgiant or late main sequence samples.
 The metallicity biases possibly hidden in the age and
 mass of the different samples are rather complicated to discuss at
 length at this point, we refer their full examination to
 Section~\ref{seccion_resultados}.

\subsection{Non-LTE effects}
\label{non_lte_effects}

  Non-LTE (Local Thermodynamic Equilibrium) effects,
  if present,
  should not constitute a bias in the comparison between groups of
  stars at the same evolutionary stage, provided that the samples are
  composed of a statistically significant number of  stars,
  showing similar properties (see previous section).
  Another issue, however, is whether non-LTE effects might 
  bias the comparison
  of samples of stars at different stages on their evolution
  (dwarf/subigant/giant), by
  affecting in bulk the abundances determination
  within a group of stars.

  Non-LTE corrections to the abundance determination increase with decreasing [Fe/H] and $\log g$, showing a strong
  dependence on the effective temperature in dwarf stars \citep[e.g.][]{2011JPhCS.328a2002B}.
  For the giant stars considered in this work, non-LTE corrections
  would be $\lesssim$ 0.1 dex, while
  for the hottest (i.e., the ``worst'' case) subgiants and late main-sequence stars, non-LTE corrections
  could be up to $\sim$ 0.1 dex \citep[][Figure~3]{2011JPhCS.328a2002B}.
  \cite{2011A&A...528A..87M} also analyzes LTE and non-LTE iron abundances for
  five stars covering a wide range of stellar parameters (T$_{\rm eff}$: 4600 - 6400 K, $\log g$: 1.60 - 4.5 dex,
  [Fe/H]: -2.7 to +0.10 dex). The authors find that
  departures from LTE do not exceed 0.1 dex for stars with solar metallicity and mildly metal-deficient stars.

  When significant departures from LTE populations in Fe~{\rm I} and Fe~{\rm II} are present,
  an LTE analysis 
  produces systematically
  underestimated gravities and metallicities
  \citep[e.g.][]{2012MNRAS.427...50L} . 
  Therefore, the comparison of
  $\log g_{\rm spec}$  and $\log g_{\rm evol}$ provides a mechanism to investigate
  whether non-LTE effects are significant or not. As discussed in
  Section~\ref{previous_work}, the standard deviation of the distribution
  $\log g_{\rm spec}$ -  $\log g_{\rm evol}$ is of the same order of magnitude
  of the uncertainties in the spectroscopic $\log g$ values. In addition, a linear fit of
  ($\log g_{\rm spec}$ -  $\log g_{\rm evol}$) with T$_{\rm eff}$ gives a slope
  consistent with zero ($\sim$ 10$^{-7}$ dex/K). The dependence with the stellar metallicity
  is more evident, although still the slope is consistent with
  zero ($\sim$ 0.05 dex/dex)\footnote{Excluding the metal-poor star BD+20 2457}.
  In other words, the good agreement between $\log g_{\rm spec}$ and $\log g_{\rm evol}$ values
  over the range of T$_{\rm eff}$ and [Fe/H] analysed in this work suggests that there are not
  significant departures from LTE.

  Considering other elements (Na, Mg, Si, ...) the only comparisons performed through this
  paper are between GWPs and GWOPs (Section ~\ref{discussion_pollution2}). Although 
  abundances for individual stars may be affected by non-LTE effects, those effects,
  if present, should not bias the comparison GWPs/GWOPs.

\section{Results}
\label{seccion_resultados}
\subsection{Analysis of the Metallicity Distributions of the Different
Samples of Stars}
\label{resultados_gigantes}


\begin{figure*}
\centering
\begin{minipage}{0.48\linewidth}
\includegraphics[angle=270,scale=0.45]{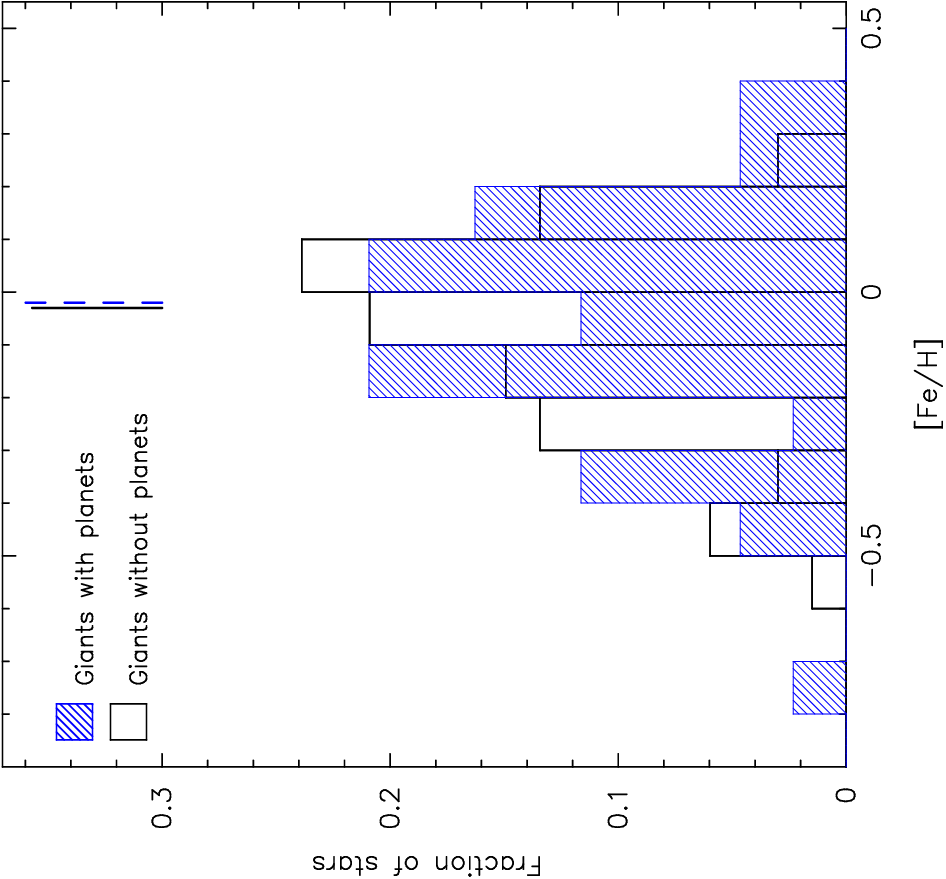}
\end{minipage}
\begin{minipage}{0.48\linewidth}
\includegraphics[angle=270,scale=0.45]{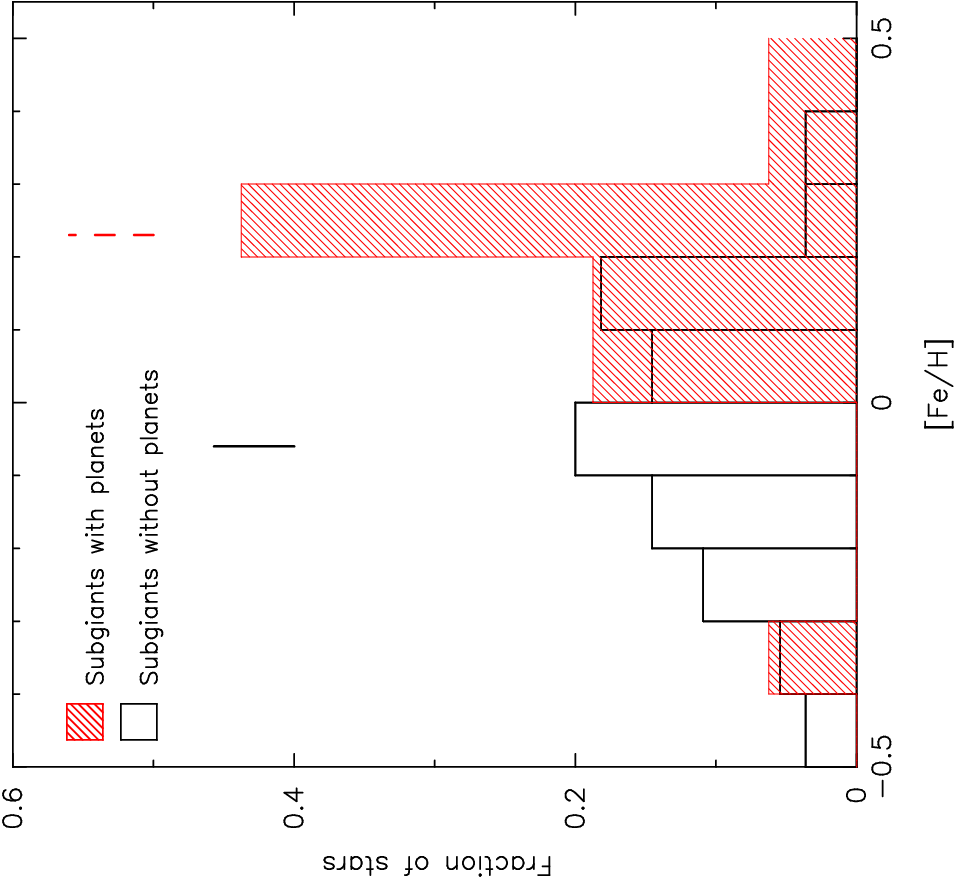}
\end{minipage}
\caption{
Left: Normalised  metallicity  distribution  of  the  GWP  sample (blue histogram)
versus  giant stars  without known planets.
Right: Normalised  metallicity  distribution  of  the  SGWP sample (red histogram)
versus  subgiant stars  without known planets.
Median values of the distributions are
shown with  vertical lines.
}
\label{histogramas}
\end{figure*}

  As mentioned before, our observations contain 67 giant stars without known planets
  and 43 giant stars with planets.
  Some statistical diagnostics for the GWOP and GWP samples are summarised
  in Table~\ref{metal_statistics} and their normalised metallicity
  distributions are shown in Figure~\ref{histogramas} (left).
  We find that both samples show similar distributions and statistical
  diagnostics. However, in order to assess if both distributions are equal from a statistical
  point of view, the standard two-sample Kolmogorov-Smirnov (K-S) test was performed.
  The maximum difference between the GWOPs and GWPs cumulative distribution
  functions is $\sim$ 0.11, and the statistical probability of both
  distributions to be drawn from the same parent distribution is
  significantly high, 87\% (n$_{\rm eff}$ $\sim$ 26).
  Therefore, we find that giant stars harbouring planets
  do not seem to follow the planet-metallicity correlation of MS
  stars. We find that giant stars with planets are not more metal rich than the
  giant stars without them. Other authors have reached the same
  conclusions before based on smaller samples of stars,
  \cite{2005PASJ...57..127S}, 
  \cite{2007A&A...473..979P}, and \cite{2008PASJ...60..781T}.

\begin{table}
\centering
\caption{[Fe/H] statistics of the stellar samples.}
\label{metal_statistics}
\begin{scriptsize}
\begin{tabular}{lcccccc}
\hline\noalign{\smallskip}
$Sample$  &  $Mean$ & $ Median$ & $Deviation$  &  $Min$&  $Max$ & $N$ \\
\hline\noalign{\smallskip}
 GWOPs    &  -0.06  & -0.03  & 0.18 & -0.50 & +0.28 & 67 \\
 GWPs     &  -0.06  & -0.02  & 0.23 & -0.79 & +0.34 & 43 \\
 SGWOPs   &  -0.06  & -0.06  & 0.22 & -0.60 & +0.35 & 55 \\
 SGWPs    &  +0.19  & +0.23  & 0.17 & -0.32 & +0.47 & 16 \\
 LMSWPs   &  +0.28  & +0.26  & 0.07 & +0.18 & +0.40 & 11 \\
\noalign{\smallskip}\hline\noalign{\smallskip}
\end{tabular}
\end{scriptsize}
\end{table}

 Figure~\ref{histogramas}, right panel, shows the normalised metallicity distribution
 of the SGWP sample and of its corresponding comparison sample (SGWOPs).
 The data suggest that the metallicity distribution of the SGWP sample is significantly
 shifted towards higher metallicities with respect to the SGWOP sample,
 a behaviour which resembles  the well known giant-planet
 metallicity correlation found in main-sequence stars
 \citep[e.g.][]{2004A&A...415.1153S,2005ApJ...622.1102F}.
 A two sample K-S test confirms that both distributions are different from a
 statistical point of view
 ($p$-value $\sim$ 10$^{-5}$, $D$ $\sim$ 0.66,
 n$_{\rm eff}$ $\sim$ 12.4).

  With the aim of completeness, a sample of main-sequence planet hosts
  (MSWPs) has been added to the discussion of the results that follows. We
  have selected those stars hosting exclusively giant planets with
  available metallicities in VF05, where we have
  removed stars with retracted or not confirmed exoplanets, as well as,
  those stars already included in our SGWP or LMSWP samples.
  In order to keep the analysis as homogeneous as possible we proceeded
  as in Section~\ref{expanding_sgwops} 
  to set the VF05 metallicities into our own metallicity scale.

\begin{figure}
\centering
\includegraphics[angle=270,scale=0.45]{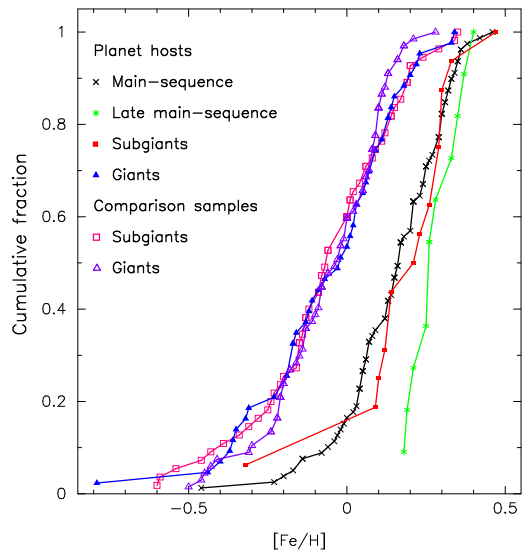}
\caption{Histogram of cumulative frequencies for the different samples studied
in this work.}
\label{distribuciones_acumuladas}
\end{figure}

 The cumulative metallicity distributions of all samples,
 Figure~\ref{distribuciones_acumuladas}, allow us to get an overall
 picture of the metallicity trends. There are a few interesting facts
 to be taken from these distributions and their statistical tests: i)
 there is not difference in the metallicities for giant stars regarding the presence or
 absence of planets; ii) the distribution of subgiant stars with
 planets is clearly separated from that of subgiants
 without planets; 
 iii) the distribution of subgiant
 stars without planets follows a similar trend than giant stars
 (with and without planets); and
 iv) more interestingly, the metallicity distribution
 of subgiant stars with planets is different from that of
 giant stars, but similar to the one of MS stars with planets.
 
 We note that the metallicity distribution of
 SGWPs seems also to be slightly shifted towards lower metallicities with respect to the
 LMSWP sample. Nevertheless, their median metallicities are quite similar
 (see Table~\ref{metal_statistics}), and both consistent with the known trends
 for main-sequence stars hosting giant-planets (see below).

 The K-S test comparing the SGWP/GWP and LMSWP/GWP samples, confirms
 that the distributions are different within a 98\% confidence level
 \footnote{$p$-value of $\sim$ 10$^{-5}$ for the SGWPs/GWPs comparison, and $\sim$ 10$^{-6}$ for the LMSWPs/GWPs
 comparison}. The K-S test reveals that the probability of LMSWPs and SGWPs to be drawn from the same parent
 distribution is low, around 0.12 ($D$=0.44, n$_{\rm eff}$ $\sim$ 6.5), although we cannot
 rule out this possibility. There is a clear outlier in the SGWP sample, namely
 HIP 36795, which is the only star in the SGWPs sample with a [Fe/H] below the solar value.
 We note that even if we do not take into account this star, the K-S probability is still low, of
 the order of 0.20. Therefore, we conclude that
 from an statistical point of view, we cannot state a difference
 between the SGWP and LMSWP samples, whilst we can affirm that, with
 the data at hand, the metal distributions of subgiant stars with
 planets are similar to the ones of late and MS stars with planets and that
 differ from those of giant stars with planets.

\subsection{Metallicity as a Function of the Stellar Mass}
\label{mass_dependence}
 
\begin{figure}[!htb]
\centering
\includegraphics[angle=270,scale=0.45]{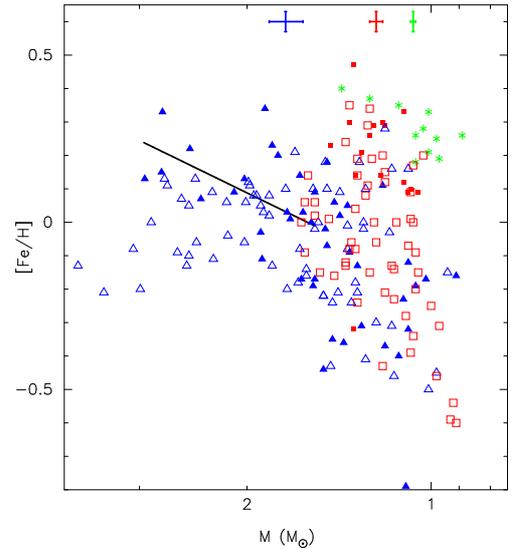}
\caption{ 
Stellar metallicity, [Fe/H], as a function of the stellar mass.
A linear fit to the data is shown to GWPs with M$_{\star}$ $>$ 1.5 M$_{\odot}$
(continuous line).
Typical uncertainties in metallicities and stellar masses are also
shown. Colours and symbols are the same as in previous figures.}
\label{metallicity_mass}
\end{figure}

 The metallicity distribution of the different samples presented in
 Figure~\ref{distribuciones_acumuladas} suggest that the
 metal-rich nature of the planet host stars tend to disappear as
 the star evolves. This could be a remarkable result that needs to be
 analyzed very carefully as there are obvious differences between the
 samples in terms of mass and age. Therefore, the data has been examined
 for correlations between mass and
 metallicity given that mass is the parameter which significantly varies between
 the giants (covering the mass range $\sim$ 1-3.8 M$_{\odot}$) and the SGWP and LMSWP samples
 (restricted to the mass range 1-1.5 M$_{\odot}$) (see also
 Table~\ref{bias_table}) and the MS samples.
 Figure~\ref{metallicity_mass} shows the [Fe/H]-Mass diagram of the
 stars analysed in this work, where the mass has been determined
 as explained in Section~\ref{previous_work}. 
 A similar plot covering the 0.8-1.2 M$_{\odot}$ mass range 
 was presented by \cite{2005ApJ...622.1102F}. Our data allow us
 to extend the plot up to 3.8 M$_{\odot}$. 
 A hint of a possible dependency of
 metallicity with stellar mass seems to appear in
 Figure~\ref{metallicity_mass} that could hinder the
 differences found for the giant stars with and without planets. Note
 that for stellar masses up to $\sim$ 1.6 M$_{\odot}$ giant stars with
 and without planets are mixed showing a lot of scatter in the graph
 and covering  the whole range of
 metallicities. However, a clear segregation in metallicity appears above the $\sim$
 1.6 M$_{\odot}$ stellar mass, the scatter in the metallicity axis is
 smaller and the giant stars with planets
 are located systematically on the metal rich part of the plot.

\begin{figure*}[!htb]
\centering
\begin{minipage}{0.48\linewidth}
\includegraphics[angle=270,scale=0.45]{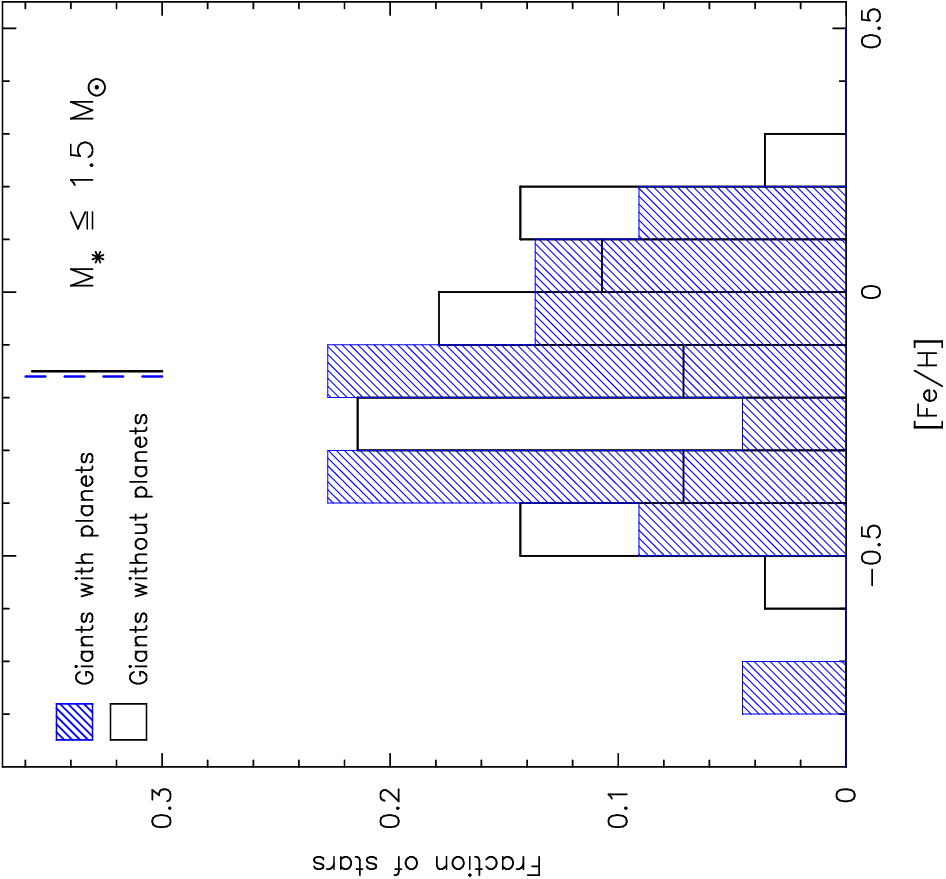}
\end{minipage}
\begin{minipage}{0.48\linewidth}
\includegraphics[angle=270,scale=0.45]{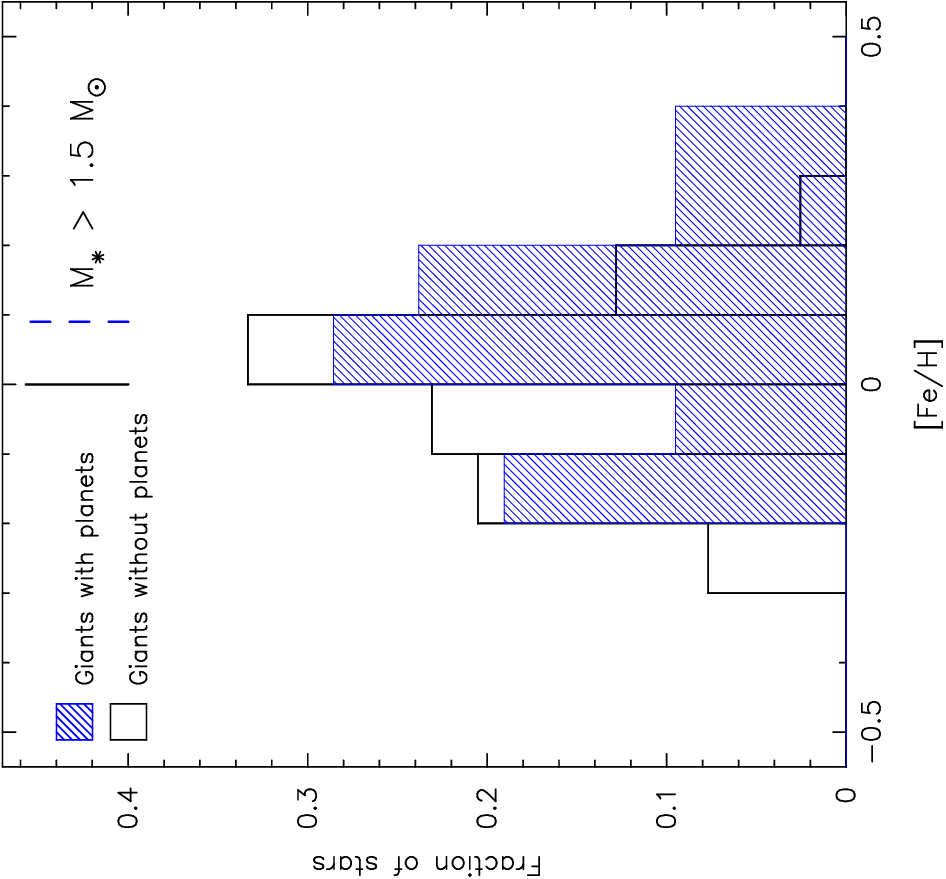}
\end{minipage}
\caption{
Left: Normalised  metallicity  distribution  of  the  GWP  sample (blue histogram)
versus  giant stars  without known planets for stars with M$_{\star}$ $\le$ 1.5 M$_{\odot}$.
Right: Normalised  metallicity  distribution  of  the  GWP sample (blue histogram)
versus  giant stars  without known planets for stars with M$_{\star}$ $>$ 1.5 M$_{\odot}$.
Median values of the distributions are
shown with  vertical lines.
}
\label{histogramas_masas}
\end{figure*}

 So, we find that for giant stars  as a whole there is no correlation between
 the presence of giant planets and the metallicity of the star, but
 within the lack of correlation there seems to be hidden a dependency with the stellar mass. 
 In the light of Figure~\ref{metallicity_mass} we have studied the metallicity distribution
 of the giant stars in the sample separated 
 according to their mass, those under 1.5 M$_{\odot}$ and
 those with larger masses. The 1.5 M$_{\odot}$ mass value has been
 chosen so that a subsample of the giants cover the same mass range as
 the subgiant sample. The histograms of the distributions are shown
 in Figure~\ref{histogramas_masas}, while some statistic diagnostics are given
 in Table~\ref{giant_statistics}. We find that the  GWPs and GWOPs samples are
 clearly separated in metallicity when only stars with M $>$1.5 M$_{\odot}$
 are considered. A K-S test shows that the probability of
 GWPs and GWOPs to be drawn from the same parent population is 
 $p$-value $\sim$ 0.70 when considering only stars with M$_{\star}$ $\le$ 1.5 M$_{\odot}$
 ($D$ $\sim$ 0.19, n$_{\rm eff}$ $\sim$ 12.3), 
 while when considering the giants with masses larger than 1.5 M$_{\odot}$, the
 K-S test probability diminishes significantly, $p$-value $\sim$ 0.05
 ($D$ $\sim$ 0.35, n$_{\rm eff}$ $\sim$ 13.7).
 
\begin{table}
\centering
\caption{[Fe/H] statistics of the sample of giant stars separated in two ranges of mass.}
\label{giant_statistics}
\begin{scriptsize}
\begin{tabular}{lcccccc}
\hline\noalign{\smallskip}
$Sample$  &  $Mean$ & $ Median$ & $Deviation$  &  $Min$&  $Max$ & $N$ \\
\hline\noalign{\smallskip}
\multicolumn{7}{c}{M$_{\star}$ $\le$ 1.5 M$_{\odot}$}\\
\hline\noalign{\smallskip}
 GWOPs    &  -0.12  & -0.15  & 0.22 & -0.50 & +0.28 & 28 \\
 GWPs     &  -0.19  & -0.16  & 0.22 & -0.79 & +0.18 & 22 \\
\hline\noalign{\smallskip}
\multicolumn{7}{c}{M$_{\star}$ $>$ 1.5 M$_{\odot}$}\\
\hline\noalign{\smallskip}
 GWOPs    &  -0.01  & +0.00  & 0.11 & -0.21 & +0.21 & 39 \\
 GWPs     &  +0.07  & +0.09  & 0.15 & -0.19 & +0.34 & 21 \\
\noalign{\smallskip}\hline\noalign{\smallskip}
\end{tabular}
\end{scriptsize}
\end{table}

 As explained in Section~\ref{possible_biases}, there are four GWPs located
 further than 200 pc with significant negative metallicities, while there are no
 similar comparison stars beyond this distance. 
 We note that these
 four stars fall in the mass domain M$_{\star}$ $\le$ 1.5 M$_{\odot}$.
  It is important to check if these low-metallicity stars are biasing
  our GWP sample in such a way that they are
  preventing us to reproduce a planet-metallicity correlation in giants
  with M$_{\star}$ $\le$ 1.5 M$_{\odot}$. If it was the case,
  removing these four stars should shift the GWP sample towards higher metallicities.
  If we repeat the K-S test for the GWPs/GWOPs samples within this
  mass domain, but removing these four stars, the $p$-value increases up to 
  roughly 80\% ($D$ $\sim$ 0.19, n$_{\rm eff}$ $\sim$ 11).
  So even excluding these four low-metallicity stars, 
  GWPs and GWOPs  in the mass domain 
  M$_{\star}$ $\le$ 1.5 M$_{\odot}$ show a similar metallicity distribution.
  Therefore, the lack of a planet-metallicity correlation in this mass-domain
  is not related to the inclusion of GWPs with low-metallicitities located 
  at larger distances.
  
  
 A search for a correlation between [Fe/H] and stellar mass has been performed for the GWP sample.
 For the giant hosts with M$_{\star}$ $>$ 1.5 M$_{\odot}$ a Spearman's correlation
 test gives a probability of correlation of the order of 99\%. 
 A linear fit to the data has been done
 and it is shown in Figure~\ref{metallicity_mass}  (continuous line).
 For the giant hosts with masses below 1.5 M$_{\odot}$ there seems to be no correlation
 between metallicity and stellar mass, the probability of a non-correlation is around 
 0.24, in other words, the correlation is not significantly different from zero.

 \cite{2005ApJ...622.1102F} found a correlation between metallicity
  and stellar mass in the 0.8-1.2 M$_{\odot}$ mass domain. The authors, however,
  note that such trend does not seem to be real (i.e., it is not
  related to the properties of the stars) but instead ``artificial'' ,
  i. e.  consequence of  stellar evolution and the colour and magnitude cuts used
  in planet search programs for targets selection.
  \cite{2010PASP..122..905J} also notices an artificial mass-metallicity correlation in a sample of 
  246 subgiants with stellar masses between 1.4-2.0 M$_{\odot}$.
  It is difficult to firmly establish if a similar effect could be the reason for the
  metallicity-mass relationship found for GWPs in the mass domain M$_{\star}$ $>$ 1.5 M$_{\odot}$,
  since the  GWP sample is composed of stars selected in different planet search programmes 
  with (probably) different criteria, sampling different regions of the HR diagram.
  The GWOP sample does not help since it is drawn from another source,
  the  \cite{2008AJ....135..209M} compilation.
  Nevertheless, most planet search programmes apply
  cuts in colours and magnitudes \citep[see e.g.][Figure~1]{2006ApJ...652.1724J}, 
  so we cannot rule out the possibility that the mass-metallicity
  relation in M$_{\star}$ $>$ 1.5 M$_{\odot}$ could be related to selection effects.


\subsection{Metallicity as a function of the Stellar Radius}
\label{discussion_pollution}

 In subgiant stars, the envelope is still cooling and expanding, in part at the expense
 of the energy being supplied by the hydrogen burning-shell, and they
 do not become fully convective until they reach the base of the red giant branch
 ascending track on the HR diagram. Giant stars, on the other hand,
 have fully convective envelopes. Therefore they both offer an unique opportunity to test the pollution
 hypothesis of planet formation. Within this scenario, high stellar
 metallicity of planet hosts is simply produced as a consequence of the accretion of
 gas depleted material on the convective zone of the star
 \citep{1997MNRAS.285..403G,1997ApJ...491L..51L}.
 Given that in this framework the metallicity would be confined to the
 convective zone in MS stars, only the external layers
 are affected. It is thus expected that the metallicity signature would be
 lost as the star evolves and the external
 metal-rich layers are gradually diluted when the convective
 zone penetrates the envelope. So late-stage accretion of material would produce
 several observables and a tendency to systematically lower
 metallicities would be expected as the star evolves from the main
 sequence to the subgiant stage and finally up the red giant phase. A way to disentangle the
 metallicity signature with evolution is exploring if there is any dependency with the
 radius of the star.

\begin{figure}[!htb]
\centering
\includegraphics[angle=270,scale=0.45]{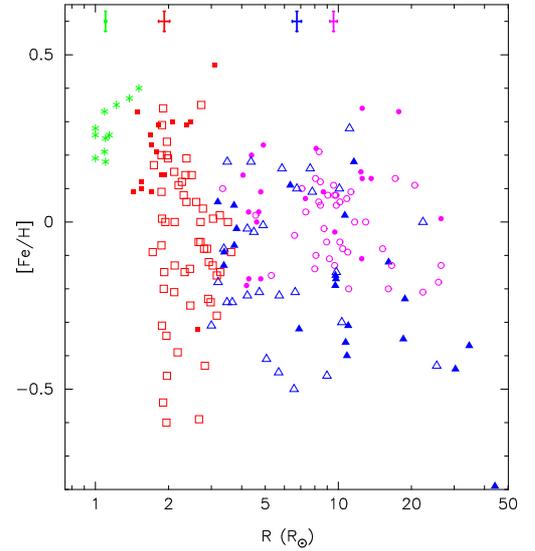}
\caption{ 
Stellar metallicity, [Fe/H], as a function of the stellar radius. Colours and
symbols are the same as in previous figures for LMSWP, SGWP, and SGWOP samples.
Giants with M$_{\star}$ $\le$ 1.5 M$_{\odot}$ are plotted in blue triangles,
while giants with M$_{\star}$ $>$ 1.5 M$_{\odot}$ are shown with
purple circles.
In both cases filled symbols indicate planet hosts.
Typical uncertainties in metallicities and stellar radius are also
shown.
}
\label{metallicity_radio}
\end{figure}

 The stellar metallicity as a function of the stellar radius for the GWP and
 GWOP samples is shown in Figure~\ref{metallicity_radio}.
 The stellar radius have been computed as explained in Section ~\ref{previous_work}.  
 Different colours
 and symbols are used for stars with masses lower than 1.5 M$_{\odot}$ and
 stars with masses greater than 1.5 M$_{\odot}$. Besides the expected
 trend towards larger radius as the stellar sample considered is more
 evolved, no other obvious trend is apparent in  Figure~\ref{metallicity_radio}.
 A very mild trend of decreasing metallicities with increasing
 stellar radius for GWPs with  M$_{\star}$ $\le$ 1.5 M$_{\odot}$ is
 doubtful as it disappears if we remove the 3 stars with the largest
 radius (94\% and  95\%  Pearson and Spearman tests respectively).

 When considering the GWPs with masses greater than
 1.5 M$_{\odot}$ only, no correlation between metallicity and radius is found. 
 The probabilities of non-correlation are $\sim$ 0.26 (Pearson's test), $\sim$ 0.14 (Spearman's test).


\subsection{Other Chemical Signatures}
\label{discussion_pollution2}

\begin{figure*}[!htb]
\centering
\begin{minipage}{0.48\linewidth}
\includegraphics[angle=270,scale=0.45]{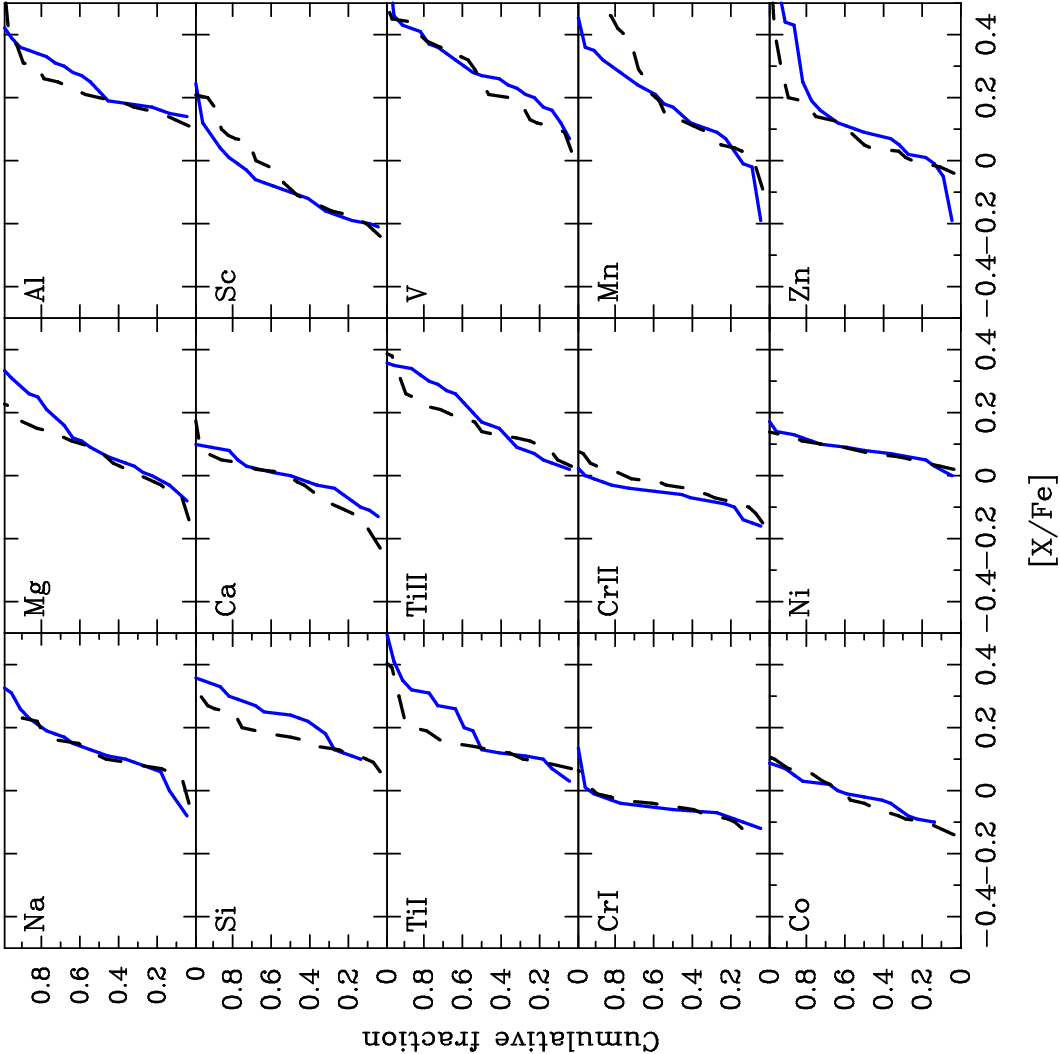}
\end{minipage}
\begin{minipage}{0.48\linewidth}
\includegraphics[angle=270,scale=0.45]{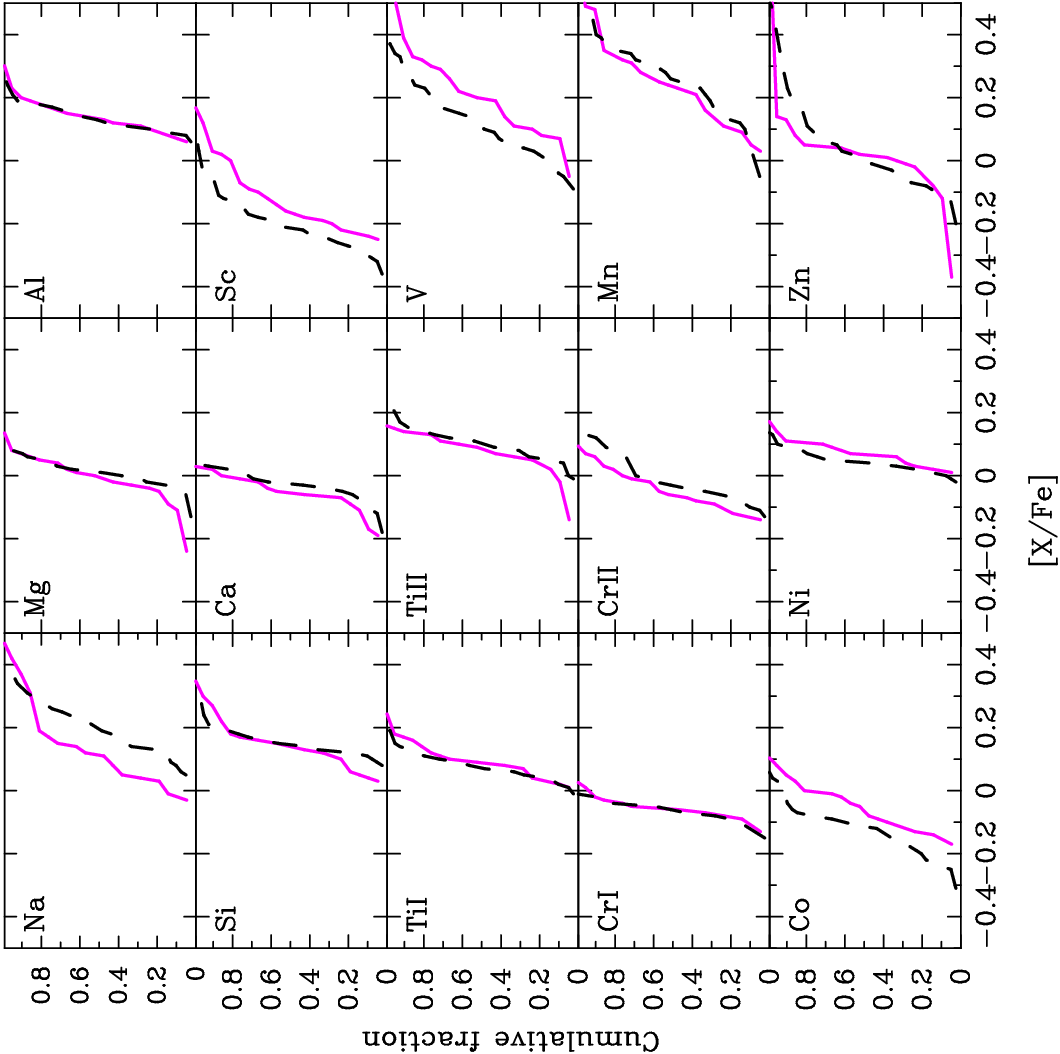}
\end{minipage}
\caption{$[X/Fe]$ cumulative fraction of
GWPs and GWOPs.
Left: Stars with M$_{\star}$ $\le$ 1.5 M$_{\odot}$, GWPs (blue continuous line)
against GWOPs (black dashed line).
Right: Stars with M$_{\star}$ $>$ 1.5 M$_{\odot}$,
GWPs (purple continuous line) against GWOPs (black dashed line).}
\label{abundancias_gigantes}
\end{figure*}

 In order to try to disclose differences in the abundances of other
 chemical elements besides Iron we show in  Figure~\ref{abundancias_gigantes}
 the cumulative distribution [X/Fe] 
 comparing the abundances [X/Fe] (where X represents Na, Mg, Al, Si, Ca, Sc, 
 Ti~{\sc i}, Ti~{\sc ii}, V, Cr~{\sc i}, Cr~{\sc ii}, Mn, Co, Ni, and Zn),
 between GWPs and GWOPs.
 On the left panel the distributions
 for giants with masses M$_{\star}$ $\le$ 1.5 M$_{\odot}$ are shown,
 while on the right panel we show the giants
 with masses M$_{\star}$ $>$ 1.5 M$_{\odot}$.
 Some statistic diagnostics are shown in Table~\ref{abund_table},
 where the results of a K-S test for each ion are also listed.

 For giants in the mass domain M$_{\star}$ $\le$ 1.5 M$_{\odot}$
 a similar behaviour between planets hosts and stars without planets
 is found. From the 15 chemical species analysed, in eight the 
 K-S probabilities are considerably high ($\ge$ 70\%), specially
 when considering Na, Ni, and Ca. In the rest, although the probabilities
 are not high, they are not significant low to state a difference
 between GWPs and GWOPs. The only remarkable exception is Si, 
 for which the GWPs distribution seems to be slightly shifted towards higher
 abundances.   

 For stars with M$_{\star}$ $>$ 1.5 M$_{\odot}$, there are significant
 differences between planets hosts and
 stars without planets
 in three species
 namely, Na (GWPs showing slightly lower abundances) and
 Co and Ni  where abundances of GWPs seem to be higher than the ones
 of GWOPs. 
 However, GWPs and GWOPs show very similar behaviours in
 Cr~{\sc i} and Al.   

  The question of whether main-sequence planet hosts show (or not)
  over-abundances of refractory elements\footnote{Elements with
  condensation temperatures near or above the condensation
  temperature of iron.} is still open
  \citep[see e.g.][and references therein]{2012A&A...543A..89A}.
  An overabundance of refractory
  elements with respect to volatiles in main-sequence planet hosts is
  considered as a possible sign of late-stage accretion,
  a tendency that is expected to disappear during the star
  evolution towards the red-giant phase.

  We find that GWPs show similar abundance patterns in all the
  elements analysed to that of GWOPs in the mass domain 
  M$_{\star}$ $\le$ 1.5 M$_{\odot}$.
  We do not know whether this is due to
  {\it i)} mixing processes which diluted the refractory enrichment previously suffered by the stars'
  progenitors; or because {\it ii)} pollution played a rather little role, so the GWPs
  progenitors never showed overabundance of refractory elements. 
  On the other hand, for masses larger than 1.5 M$_{\odot}$,
  GWPs and GWOPs show differences in some elements, specially
  Na, Co, and Ni.


\begin{table}[!htb]
\centering
\caption{Comparison between the elemental abundances of GWPs and GWOPs.}
\label{abund_table}
\begin{scriptsize}
\begin{tabular}{lcccccc}
\hline\noalign{\smallskip}
               &  \multicolumn{6}{c}{M$_{\star}$ $\le$ 1.5 M$_{\odot}$}  \\
               &  \multicolumn{6}{c}{\hrulefill}                         \\
               &  \multicolumn{2}{c}{\textbf{GWPs}}   & \multicolumn{2}{c}{\textbf{GWOPs }} & \multicolumn{2}{c}{\bf K-S test$^{\dag}$}  \\
 $[X/Fe]$      &  \multicolumn{2}{c}{\hrulefill}      & \multicolumn{2}{c}{\hrulefill}      & \multicolumn{2}{c}{\hrulefill}             \\
               &  Median       &  Deviation           & Median      & Deviation             & $p$-value     &  $D$                       \\
\hline\noalign{\smallskip}
 Na          &  0.12 & 0.10 &  0.12 & 0.08 & 0.96 & 0.14 \\
 Mg          &  0.09 & 0.12 &  0.09 & 0.09 & 0.39 & 0.25 \\
 Al          &  0.24 & 0.09 &  0.21 & 0.10 & 0.42 & 0.24 \\
 Si          &  0.25 & 0.09 &  0.18 & 0.06 & 0.04 & 0.39 \\

 Ca          &  0.00 & 0.07 &  0.01 & 0.09 & 0.90 & 0.16 \\
 Sc          & -0.10 & 0.12 & -0.07 & 0.14 & 0.75 & 0.19 \\ 
 Ti~{\sc i}  &  0.16 & 0.13 &  0.14 & 0.08 & 0.17 & 0.30 \\
 Ti~{\sc ii} &  0.19 & 0.12 &  0.16 & 0.09 & 0.30 & 0.27 \\
 V           &  0.28 & 0.14 &  0.29 & 0.14 & 0.71 & 0.19 \\
 Cr~{\sc i}  & -0.06 & 0.06 & -0.05 & 0.05 & 0.86 & 0.17 \\
 Cr~{\sc ii} & -0.05 & 0.05 & -0.03 & 0.05 & 0.18 & 0.30 \\
 Mn          &  0.18 & 0.15 &  0.14 & 0.20 & 0.26 & 0.28 \\ 
 Co          & -0.02 & 0.06 & -0.04 & 0.08 & 0.87 & 0.16 \\   
 Ni          &  0.09 & 0.04 &  0.09 & 0.03 & 0.96 & 0.14 \\
 Zn          &  0.10 & 0.20 &  0.07 & 0.17 & 0.71 & 0.19 \\  
\hline\noalign{\smallskip}
\multicolumn{7}{l}{$^{\dag}$ n$_{\rm eff}$ $\sim$ 12.3}\\
\noalign{\smallskip}\hline\noalign{\smallskip}
               &  \multicolumn{6}{c}{M$_{\star}$ $>$ 1.5 M$_{\odot}$}  \\
               &  \multicolumn{6}{c}{\hrulefill}                         \\
               &  \multicolumn{2}{c}{\textbf{GWPs}}   & \multicolumn{2}{c}{\textbf{GWOPs }}   & \multicolumn{2}{c}{\bf K-S test$^{\ddag}$} \\
 $[X/Fe]$      &  \multicolumn{2}{c}{\hrulefill}      & \multicolumn{2}{c}{\hrulefill}        & \multicolumn{2}{c}{\hrulefill}             \\
               &  Median       &  Deviation           & Median      & Deviation               & $p$-value     &  $D$                       \\
\hline\noalign{\smallskip}
 Na          & 0.12  & 0.14 & 0.21  & 0.09 & 0.01 & 0.42 \\
 Mg          & 0.00  & 0.08 & 0.02  & 0.05 & 0.73 & 0.18 \\
 Al          & 0.14  & 0.06 & 0.13  & 0.05 & 0.95 & 0.14 \\
 Si          & 0.15  & 0.08 & 0.15  & 0.05 & 0.66 & 0.19 \\
 Ca          & -0.05 & 0.06 & -0.02 & 0.05 & 0.06 & 0.34 \\
 Sc          & -0.16 & 0.12 & -0.21 & 0.10 & 0.11 & 0.32 \\ 
 Ti~{\sc i}  & 0.09  & 0.06 & 0.08  & 0.05 & 0.59 & 0.20 \\
 Ti~{\sc ii} & 0.09  & 0.07 & 0.11  & 0.05 & 0.70 & 0.18 \\
 V           & 0.20  & 0.15 & 0.11  & 0.12 & 0.04 & 0.36 \\
 Cr~{\sc i}  & -0.05 & 0.03 & -0.06 & 0.03 & 0.98 & 0.12 \\
 Cr~{\sc ii} & -0.06 & 0.07 & -0.03 & 0.08 & 0.24 & 0.27 \\
 Mn          & 0.24  & 0.15 & 0.26  & 0.15 & 0.84 & 0.16 \\
 Co          & -0.05 & 0.08 & -0.11 & 0.08 & 0.02 & 0.40 \\
 Ni          & 0.07  & 0.04 & 0.05  & 0.04 & 0.01 & 0.41 \\
 Zn          & 0.02  & 0.22 & 0.02  & 0.16 & 0.49 & 0.22 \\
\hline\noalign{\smallskip}
\multicolumn{7}{l}{$^{\ddag}$ n$_{\rm eff}$ $\sim$ 13.7}\\
\noalign{\smallskip}\hline\noalign{\smallskip}
\end{tabular}
\end{scriptsize}
\end{table}

\subsection{Age-Metallicity Relation}
\label{seccion_age_trends}

\begin{figure}
\centering
\includegraphics[angle=270,scale=0.45]{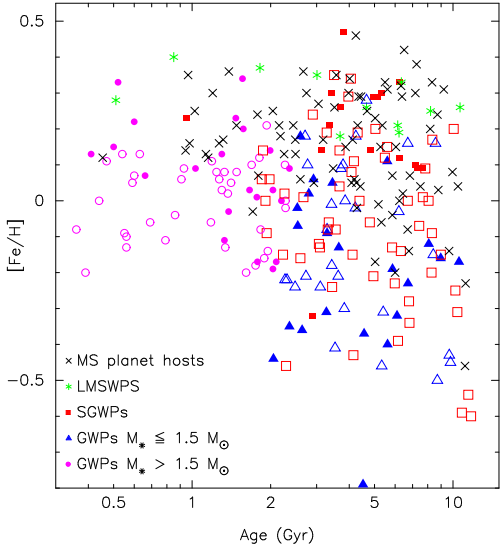}
\caption{Age-metallicity relation for the different samples studied in this work.
Open symbols indicate the corresponding comparison samples.}
\label{edad_metalicidad}
\end{figure}

 In the light of the metallicity trends with stellar mass found within the giant stars sample 
 it is reasonable to explore a possible age-metallicity relation.
 In Figure~\ref{edad_metalicidad} we show the stellar age versus its metallicity of the 
 different samples analyzed in this work. 
 For comparison, main-sequence hosts from VF05 are overplotted with VF05 [Fe/H] values
 set into our metallicity scale as explained in Section~\ref{expanding_sgwops}. 
 In addition, the stellar ages of these stars have been recomputed using the  methodology 
 followed in this work (Section~\ref{previous_work}).  
 Two clear trends can be identified  in Figure~\ref{edad_metalicidad}. To the left of the 
 plot are located the giant stars with masses M$_{\star}$ $>$ 1.5 M$_{\odot}$ stars,
 and to the right of the plot all the other stars studied. The plot shows the expected 
 trend in metallicity with stellar ages. As the population is older the metallicity have a 
 tendency to show a larger spread in values.

\subsection{Trends with the planetary properties}
\label{seccion_planetary_trends}

\begin{figure*}[!htb]
\centering
\begin{minipage}{0.48\linewidth}
\includegraphics[angle=270,scale=0.45]{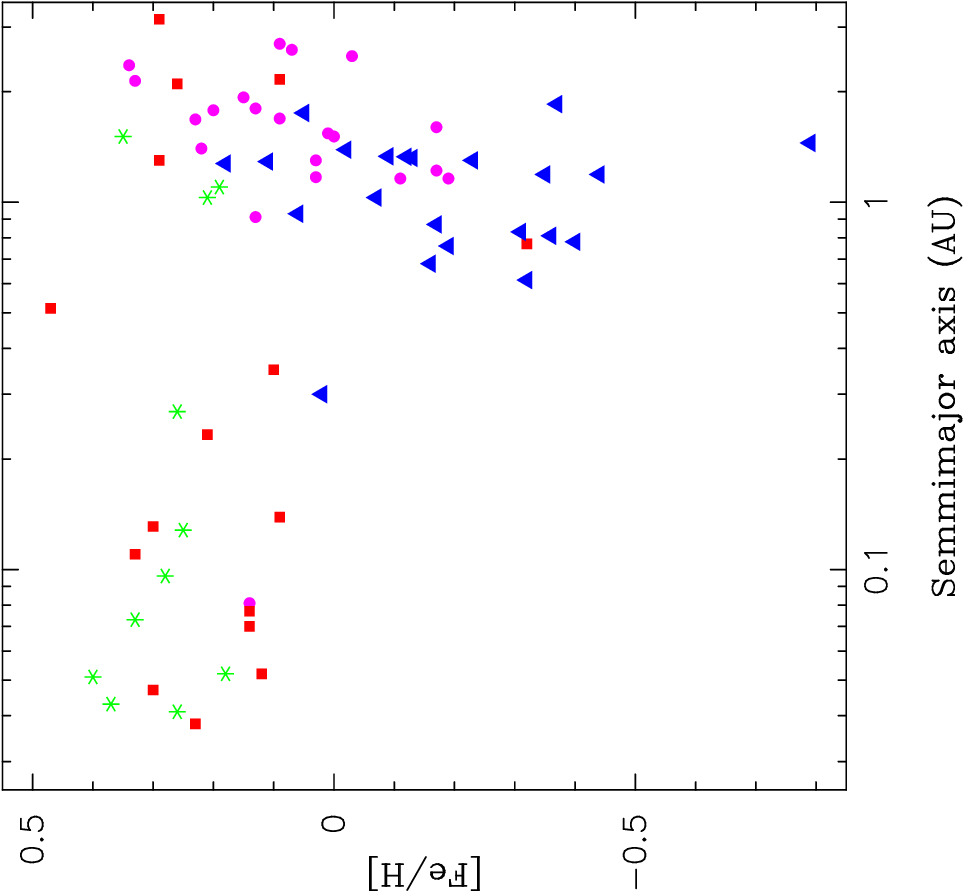}
\end{minipage}
\begin{minipage}{0.48\linewidth}
\includegraphics[angle=270,scale=0.45]{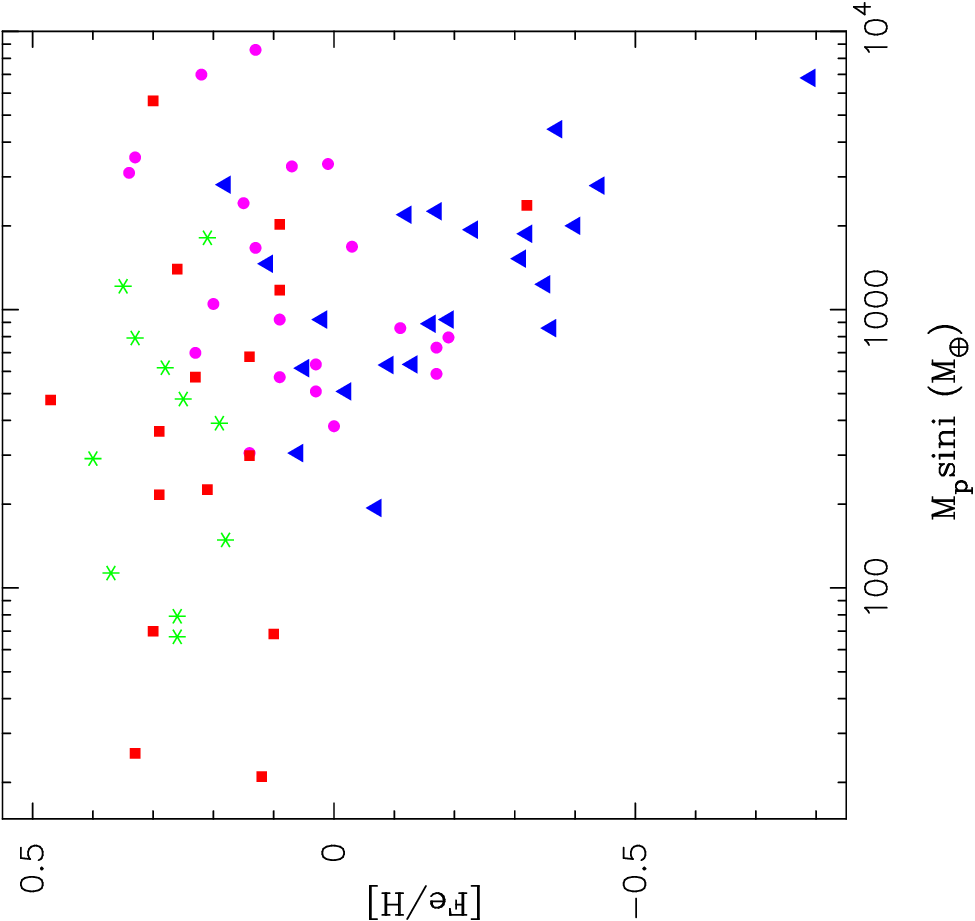}
\end{minipage}
\caption{Left: Stellar metallicity as a function of the semimajor axis of the innermost
 planet. Right: Stellar metallicity as a function of the mass of the most massive
 planet. GWPs with M$_{\star}$ $>$ 1.5 M$_{\odot}$ are plotted with purple circles, 
 GWPs less massive than 1.5 M$_{\odot}$  in blue triangles, SGWPs as red  squares, while LMSWPs are plotted
 as green asterisks.
}
\label{planetary_plots1}
\end{figure*}

Studies around MS stars have revealed that the metal signature on the star seems 
to influence the maximum mass of the planet that can be formed \citep{2011arXiv1109.2497M}. 
It has been shown that the 
planet deficiency at small orbital distances found around red giant stars \citep[see also][]{2007ApJ...665..785J,2008PASJ...60.1317S,2009ApJ...693.1084W} can be explained by
tidal interactions in the star-planet system as the star
evolves off the main-sequence, which can lead to variations
in the planetary orbits and to the engulfment of close-in planets
\citep{2009ApJ...705L..81V}. The planet accretion process can lead to a  transfer of angular momentum 
to the stellar envelope which ultimately can spin up the star and even indirectly modify its chemical abundances. 
Possible evidence of this process has been recently found \citep[see e.g.][]{2012ApJ...754L..15A}.
Furthermore,  \cite{2012ApJ...757..109C} analysed 
a sample of slow and rapid RGB rotators and found lithium enrichment on the rapid rotators 
consistent with planet accretion onto the stellar envelope. 

In order to disclose any possible trends on the planet properties among
the stellar samples studied in this work we show on the left panel of Figure~\ref{planetary_plots1}
the stellar metallicity versus the orbital distance of the planet and on the right panel 
the stellar metallicity as a function of the mass of the more massive planet
\footnote{
M$_{\rm p}\sin i$, with the exceptions of the planets orbiting around the stars GSC 2883 -01687,
HIP 80838, TrES-4,  and HAT-P-7, detected by transits.
}.
The planets on our sample follow the general trend mentioned above, that is,
nearly all planets orbiting GWPs are cool distant (a $>$ 0.5 AU)
gaseous jupiters, with the only exception of HIP 57820 (which hosts a close-in
Jupiter at a $\sim$ 0.08 AU 
and HIP 114855 (a = 0.3 AU).  Regarding the planet-mass metallicity relation (right panel)
there seems to be a trend of decreasing metallicities as we move towards higher planetary
masses. A Spearman correlation test provides a likelihood of correlation of
96\%. 
This appears to be in contradiction with the known trends of main-sequence
FGK hosts 
in which a positive correlation between the metallicity of the host star 
and the mass of its most massive planets is found
\citep[][]{2011arXiv1109.2497M,2011A&A...533A.141S}. 
  A closer inspection of the metallicity-planetary mass plane reveals that
 this general tendency is due to the GWPs stars less massive than 1.5 M$_{\odot}$.
 Considering only these stars, the likelihood of a correlation  is 
 $\sim$ 99\% (Spearman's test). 
 On the other hand, there is no obvious correlation when considering
 the other samples. The behaviour of LMSWPs and SGWPs is more or less flat,
 while for GWPs with M$_{\star}$ $>$ 1.5 M$_{\odot}$ there seems to be a hint
 of increasing metallicities with increasing planetary masses
 (a Spearman's test gives a probability of correlation of $\sim$ 91\%).
 In other words, subgiants and high-mass giants reproduce the known
 trends for main-sequence hosts, while giants in the low-mass domain
 show a behaviour which is hard to understand.

  Next, we explore the planet properties among the different samples.
  It has been suggested that giant stars host more-massive planets than
  main-sequence hosts \citep[e.g.][]{2007ApJ...665..785J,2007A&A...472..657L}.
  A comparison of the cumulative frequency of planet mass\footnote{
  We take as reference the innermost planet in multiple systems since
  radial velocity surveys are more sensitive to close-in, massive planets.
  }
  between our sample of
  GWPs and main-sequence stars from VF05
  hosting exclusively giant planets
  reveals that the distribution of the former is 
  clearly shifted towards higher masses. While main-sequence hosts spread a planetary
  mass range from 0.1 to 18 M$_{\rm Jup}$ with a median value of 1.9 M$_{\rm Jup}$,
  our sample of GWPs covers from 0.6 to 22 M$_{\rm Jup}$ with a median value of 3.3 M$_{\rm Jup}$.
  A K-S test shows that both distributions are statistically different within a confidence level
  of 98\% ($p$-value $\sim$ 10$^{-3}$). This result should be interpreted very carefully, since 
  the larger levels of jitter in evolved stars might prevent the detection of lower mass planets 
  shifting the planet mass distribution towards larger values. 
  No obvious segregation either in mass nor in orbital distance is found among the planets 
  orbiting giant stars with different masses.

  Regarding multiplicity, we find a rate of multi-planet systems in GWPs 
  of the order of 12\% which is in agreement
  with the 14\% multiple confirmed planetary systems given by 
  \cite{2009ApJ...693.1084W} although it could be 28\% or higher if those cases with
  significant evidence of being multiple are included. Finally, no correlation
  between the stellar metallicity  and the planet's eccentricity
  was found, although as pointed out by \cite{2008ApJ...675..784J} we note that
  the eccentricity distribution of GWPs seems to be shifted towards lower values
  than the eccentricity distribution of main-sequence hosts. 
  While the median eccentricity in GWPs
  is 0.15, in main-sequence hosts is around 0.25. 
  A K-S test reveals both distributions to be different 
  ($p$-value $\sim$ 10$^{-3}$).

  Planets in the SGWP sample are predominantly cool, although around 30\% of the stars host
  a hot-Jupiter at a distance closer than 0.1 AU. In addition, two of our stars in the
  SGWP sample host low-mass planets (M$_{\rm p}$ $<$ 30 M$_{\oplus}$), namely HIP 94256, and
  HIP 115100. When considering the LMSWP sample, roughly 50\% of the stars harbour at least
  one hot-Jupiter, while the other 50\% only host cool distant planets. HIP 98767 hosts two
  planets, being the innermost one a low-mass planet. 

\section{Discussion}        
\label{seccion_discussion}  

 As pointed out in Section~\ref{resultados_gigantes},
 we find that the metal distribution of subgiant stars with planets is 
 clearly separated 
 from that of subgiants without planets, and that it is similar 
 to the one of MS stars with planets. 
 Considering the whole
 sample of giant stars (i.e. without mass segregation) we do not find a difference in the metal 
 distribution of giant stars that host planets
  when compared with giant stars where no planetary systems have been detected. 
 While the metallicity distribution of the subgiants 
 fit well within the current paradigm of planet formation, the giant stars results are harder 
 to understand within this context.

 One could argue that the metallicity signature  
 of planet formation disappears at the moment the star evolves 
 into red giant branch. The M$_{\rm bol}$ criterion chosen to separate the subgiant from the giant sample 
 physically reflects the time at which the star becomes fully convective. At this point, 
 three lines of arguments could be followed: i) there was not metal difference between 
 the stars bearing planets and stars with no planets in this sample of giant stars, ii) 
 there was a different metallicity but has been lost, iii) the sample is biased in such a way 
 that prevent us from seeing any metallicity difference. 


\subsection*{Can massive proto-planetary disks explain the observed trends?}        
\label{coreacrretion}  
  Let´s explore the first possibility, i. e. that the giant stars
  represent a different stellar population in which a metal rich
  environment is not required for planet formation. 
  The red giant stars that constitute our GWP sample are the result of the evolution of early-type
  main-sequence dwarfs. If we go back in time on the evolutionary tracks, the stars in the GWP sample 
  are the result of the evolution of main-sequence dwarfs with effective temperatures in the
  range 5500 - 12000 K, (spectral-types G5V-B8V, and stellar masses between 0.9 and 4 M$_{\odot}$). 
  On the other hand, SGWPs come mainly from G5V-F0V (M$_{\star}$ between 0.9 and 1.6 M$_{\odot}$), while 
  stars in the LMSWP sample come from less massive stars with spectral types in the range  K2V-F2V.
  It is therefore natural to ask whether the observed differences in the metallicity distribution 
  of the different samples are related to the different mass distributions of the star's
  progenitors in the main-sequence \citep[e.g.][]{2010ApJ...725..721G}.
  In principle, high-mass stars are likely to harbour more massive protoplanetary disks 
  \citep[e.g.][see also Fig.~5 in \citealt{2011ARA&A..49...67W}]{2000prpl.conf..559N}.
  Observations of H$_{\alpha}$ EWs in young, low-mass objects suggest that the
  mass accretion rate scales approximately with the square of the stellar mass
  \citep{2003ApJ...592..266M,2004A&A...424..603N,2011A&A...535A..99M,2012A&A...543A..59M},
  a result which can be reproduced
  assuming that a relationship between the disk mass and the central star mass
  on the form  M$_{\rm disk}$ $\propto$ M$_{\star}^{1.2}$ holds \citep{2011A&A...526A..63A}.

  According to recent simulations of planet population synthesis
  \citep{2011A&A...526A..63A,2012A&A...541A..97M}, protoplanetary
  disk masses play a significant
  role in planet formation. In particular, it is shown that giant planet formation can occur
  in low-metallicity (low dust-to gas ratio) but high-mass protoplanetary disks.
  The metallicity effect depends on the mass of the disc, being the minimum metallicity
  required to form a massive planet correspondingly lower for massive stars than for low-mass stars.
  In this scenario, the fact that GWPs do not show the metal-rich signature, could be explained by the more massive 
  protoplanetary disks of their progenitors. 
  However, several difficulties arise.

  One of the consequences of the protoplanetary disk mass on planet formation
  is that planets orbiting massive giant stars should be more massive than planets around dwarf
  stars. However, as already mentioned in Section~\ref{seccion_planetary_trends}, one should be careful in the
  comparison between planets around giants and MS stars, given that the detections are affected
  from biases introduced by the star. The samples of giant stars with different masses
  are however suitable for this comparison and we
  find that there is not obvious difference on the minimum mass of the planet found between the low
  mass and the high mass giant stars.

  Second, core-accretion models are  not able to predict the presence of very massive companions
  around very-low metallicity stars, and moreover around stars that did not suppose to have a
  massive disk to begin with (giants with M$_{\star}$ $\le$ 1.5 M$_{\odot}$). Although those
  planets are rare, we note that there are some of them in our GWP sample, such as the two companions
  around BD+20 2457 ([Fe/H]=-0.79 dex, planets of 21.4 M$_{\rm Jup}$ at 1.5 AU, and 12.5  M$_{\rm Jup}$
  at 2.0 AU), the planet around $\gamma^{1}$ Leo ([Fe/H]=-0.44 dex, planet of $\sim$ 8 M$_{\rm Jup}$ at 1.2 AU),
  or the one orbiting HD 13189 ([Fe/H]=-0.37 dex, planet of $\sim$ 14 M$_{\rm Jup}$ at 2 AU).
  This is because the time needed to form a core big enough to start a runaway accretion of gas
  is so long that by that time the gas has already been significantly depleted.

  But the most intriguing point is the mass-segregation found for GWPs (Section~\ref{mass_dependence}).
  While the metallicity distribution of GWPs in the mass-domain M$_{\star}$ $>$ 1.5 M$_{\odot}$
  is shifted towards higher metallicities with respect to a similar sample of giants without
  planets, GWPs in the mass range M$_{\star}$ $\le$ 1.5 M$_{\odot}$ do not show the metal
  signature of the presence of planets.  
  This is a puzzling result, at least in two ways.
  First, if planet formation can occur in low-metal, but high protoplanetary disk masses environments,
  a population of massive giant stars with low metallicities hosting planets might be expected. 
  Our observations show somehow the opposite, massive (M$_{\star}$ $>$ 1.5 M$_{\odot}$) giant stars
  with planets show high-metallicities. 

  Second, it is worth to note that the sample of less massive (M$_{\star}$ $\le$ 1.5 M$_{\odot}$) giant stars
  with planets covers the same mass range as the MS progenitors
  and subgiant stars where the metal signature has been observed.
  In other words, the protoplanetary disks of GWPs and M$_{\star}$ $\le$ 1.5 M$_{\odot}$
  are not massive, and thus, there is nothing to help planet formation at low metallicities.
  Furthermore, there is no age difference between this sample of
  stars and the MS or subgiants stars. Thus, the fact that giant stars
  with M$_{\star}$ $\le$ 1.5 M$_{\odot}$
  and planets are not more metal rich is hard to understand as it is in apparent
  contradiction with the trademark of the core-accretion model.

\subsection*{Can the metallicity signature  be erased as the star evolves?}
\label{pollution}
  If we accept the possibility that giant stars do not favour the
  existence of a metal poor
  environment for planet formation, then we have to explore the option
  that the metallicity signature was present at the time the planet
  was formed but then disappeared as a consequence of the evolution of
  the star. 
 \citet[][see also \citealt{1996Natur.380..606L}]{1997MNRAS.285..403G}
 explain the metal content of planet host stars as a consequence of 
 the accretion of gas depleted material in the stellar surface, the so-called
 pollution scenario. In this scenario, only the external layers of the
 stars are affected, and as the star evolves, the external metal-rich
 layers are gradually diluted as the convective zone of the star grows.

 Our data does not support evidence of pollution. If the metal-rich signature
 was limited to the convective envelope of the stars, subgiants with planets should show lower
 metallicities than main-sequence hosts. We find the opposite, with SGWP and main-sequence
 planet host samples  showing the same chemical signature (Section~\ref{resultados_gigantes}). 
 Furthermore, the different metallicity behaviour of GWPs depending on their masses is 
 again difficult to understand in the pollution scenario. There
 is no physical reason why the metal-rich nature of the star would be
 lost due to convection only  for giant stars with M$_{\star}$ $\le$
 1.5 M$_{\odot}$, remaining for giants with M$_{\star}$ $>$ 1.5 M$_{\odot}$. The metallicity-stellar
 radius relation does not shed any light into this issue as M$_{\star}$ $\le$ 1.5 M$_{\odot}$ giant
 stars cover the same range on stellar radius as the M$_{\star}$ $>$ 1.5 M$_{\odot}$ stars. 
 
\subsection*{Is our sample biased?}
\label{other}

Other lines of arguments such as the one suggested by \cite{2009ApJ...698L...1H} in which
 the observed correlation between the presence of gas-giant planets and enhanced
 stellar metallicity observed in main-sequence planet hosts, might
 be related to a possible inner disk origin of these stars does not fit the data either.
 In this scenario, the observed metallicity distribution of GWPs would be shifted towards
 lower metallicities with respect to the one of main-sequence hosts, just simply because
 the GWP sample contains stars younger than the dwarf sample and, therefore, less contaminated
 by radial mixing. Nevertheless, according to this scenario, giant stars  with planets
 and high-masses (M$_{\star}$ $>$ 1.5 M$_{\odot}$) should not be metal-rich. 
 
 The possible biases affecting our sample have been explored in the
 paper. We have found no biases in age, mass, population,
 or distance that could explain our results. However, an option that
 we cannot exclude is the risk of the sample size being small. To fix
 this issue we will have to wait for more planet discoveries to take place.

\subsection{Mass segregation and previous results}
\label{mass_segregation}


 We should finally discuss how the mass-segregation found for GWPs in this work compares with
 previous results on evolved stars with planets. It is worth to note that
 most of the stars included in previous works are in the mass-domain
 M$_{\star}$ $\le$ 1.5 M$_{\odot}$, where the metal-rich signature of planet hosts
 is lost. 
 Specifically, the number of low-mass giants in each work are: 
 2 out the 4 stars analyzed in \cite{2005PASJ...57..127S};
 1/1 in \cite{2005ApJ...632L.131S};
 7/10 in \cite{2007A&A...473..979P}; and  7/16 in \cite{2010ApJ...725..721G}. 
 However, among the 20 giants included in \cite{2007A&A...475.1003H}, 11 are
 high-mass (M$_{\star}$ $>$ 1.5 M$_{\odot}$) giants. The fact that GWPs
 in the high-mass domain show-metal enrichment, while less-massive giants do not,
 could explain the disagreement between \cite{2007A&A...475.1003H} and 
 other works. 
 We note, however, that \citet{2008PASJ...60..781T} do not found metal-enrichment
 in GWPs, despite the fact of 7 out of the 10 stars analyzed in that work are in the 
 high-mass domain.

\section{Conclusions}
\label{conclusions}

  Evolved stars (subgiants and red giants) with planets constitute valuable tools
  to set constrains into our understanding on how planetary systems do form and
  evolve. Nowadays, an increasing effort is being applied in searching for planetary
  companions around this kind of stars. In addition,
  the properties of evolved stars with planets, and also the properties of the planets
  found around these stars seem to be different from which we already know for main-sequence
  planet hosts. 
   In this work, we perform an analysis of the stellar properties
  and elemental abundances of a large sample of evolved stars. 
  Although data from the literature has been used to expand the SGWOP sample, our analysis
  has, to our best knowledge, the best combination between homogeneity and sample size discussed so far. 
  In addition, a detailed analysis of the stellar samples
  properties is performed in order to avoid any bias which could affect our results.
 

  We find that, unlike the case of main-sequence hosts, planets around giant stars are not
  preferentially found around metal-rich stars when the whole sample of giant stars 
  is analyzed. The metallicity distribution of
  GWPs is clearly shifted towards lower values in comparison with SGWPs and
  LMSWPs. Taken into account the homogeneous procedure followed
  in this work, and the fact that we are mainly dealing with solar-type stars,
  we state that the differences in the metallicity distributions are real and  
  not due to non-LTE effects. 

  Subgiant stars show the same metal trends with planets as MS stars and also have 
  a similar mass range. In an attempt to understand if the more massive stars within 
  the giant sample are  shifting the metal distribution of the GWP 
  towards lower metallicities, we segregated the giant stars in two mass bins,  
  M$_{\star}$ $\le$ 1.5 M$_{\odot}$ and  M$_{\star} >$ 1.5 M$_{\odot}$. It is shown
  for the first time that the metallicity 
  distribution of the more massive giant stars with planets is 
  shifted towards higher metallicities, as it is the one for the MS and subgiant stars.


  The metal signature of the presence of planets is lost, however, for stars in the
  M$_{\star}$ $\le$ 1.5 M$_{\odot}$ range, a fact which is difficult to understand
  with current models of planet formation. 
  These stars show a similar range of stellar
  parameters
  than subgiant and main-sequence
  planet hosts but, do not show the metal-enrichment signature.
  In particular, giants with M$_{\star}$ $\le$ 1.5 M$_{\odot}$ show
  a similar age distribution than subgiants and main-sequence hosts,
  ruling out radial mixing as a possible explanation of their metallicity
  distribution. 
  Since they also show similar masses, 
  a planet formation scenario in which low-metallicity environments are compensated
  by higher-mass protoplanetary disks, can be also discarded.
  Taken into account that no bias that could affect the metallicity distribution of low-mass giant 
  hosts has been identified, the only explanation 
  points towards a non-primordial origin of the metallicity-gas giant
  planet relationship. 
  We have, however, not found clear evidence of pollution and
  furthermore, what is more intriguing,
  why convection should play a role erasing the metal signature for
  giants in the mass domain M$_{\star}$ $\le$ 1.5 M$_{\odot}$, and 
  not for giants with  M$_{\star} >$ 1.5 M$_{\odot}$?.

  Additional differences between giants with masses 
  $\le$ 1.5 M$_{\odot}$ and more massive giants have been found
  when analysing the abundance patterns of different elements.
  While in the case of the less-massive giants, planet hosts and
  non-planet hosts show similar abundance patters, 
  in the more massive stars there are differences in some elements between
  stars hosting planets and stars without known planets,
  specially in the cases of Na, Co, and Ni abundances. 

  Finally, we note that
  planets around evolved stars show some peculiarities with respect 
  to the planets orbiting around main-sequence stars, like a
  lack of close-in planets or higher masses and eccentricities.
  The data also suggest a decreasing trend between the stellar metallicity
  and the mass of the most massive planet.

\begin{acknowledgements}

  This work was supported by the Spanish Ministerio de Ciencia e Innovaci\'on (MICINN),
  Plan Nacional de Astronom\'ia y Astrof\'isica, under grant 
  \emph{AYA2010-20630} and \emph{AYA2011-26202}.
  J.M. acknowledges support from the Universidad Aut\'onoma de Madrid
  (Department of Theoretical Physics). E. V. 
  also acknowledges the support provided by the Marie Curie 
  grant \emph{FP7-People-RG268111}. The authors would like to thank
  Robert. L. Kurucz, 
  Sergio Sousa, Yoichi Takeda, and L\'eo Girardi,  
  for making their codes
  publicly available. Jean Schneider
  is also acknowledged for maintaining the Extrasolar Planets Encyclopedia. 
  We sincerely appreciate the careful
  reading of the manuscript and the constructive comments of an anonymous
  referee.

\end{acknowledgements}


\bibliographystyle{aa}
\bibliography{gigantes.bib}


\Online
\section*{Online material}
\label{tables}

  Results produced in the framework of this work 
  are only available 
  in the electronic version of the corresponding paper or
  at the CDS via anonymous ftp to cdsarc.u-strasbg.fr (130.79.128.5)
  or via {\tt http://cdsweb.u-strasbg.fr/cgi-bin/qcat?J/A+A/}
 
  Table~\ref{parameters_table} contains:
  HIP number (column 1); HD number (column 2); 
  effective temperature in Kelvins (column 3); logarithm of the surface gravity in cms$^{\rm -2}$ (column 5);
  microturbulent velocity in km$s^{\rm -1}$ (column 5); final metallicity in dex (column 6);
  mean iron abundance derived from Fe~{\sc I} lines (column 7) in the usual scale
  ($ A(Fe) = \log[(N_{Fe}/N_{H}) + 12$]); number of Fe~{\sc I} lines used (column 8);
  mean iron abundance derived from Fe~{\sc II} lines (column 9);
  number of Fe~{\sc II} lines used (column 10); and spectrograph (column 11).
  Each measured quantity is accompanied by its corresponding uncertainty.

  Table~\ref{evolutionary_table} gives:
  HIP number (column 1); visual extinction in magnitudes (column 2);
  $\log$(L$_{\star}$/L$_{\odot}$) (column 3); photometric effective temperature
  in Kelvins (column 4);  evolutionary values of the
  logarithm of the surface gravity in cms$^{\rm -2}$ (column 5);
  stellar age in Gyr (column 6); stellar mass in solar units (column 7);
  and stellar radius in solar units (column 8). 
  Each quantity is accompanied by its corresponding uncertainty.

  Table~\ref{tabla_abundancias} gives the
  abundances of Na, Mg, Al, Si, Ca, Ti~{\sc I}, Ti~{\sc II},
  V, Cr~{\sc I},
  Cr~{\sc II}
  Mn, Co, Ni,
  and Zn (Table~10). 
  They are expressed relative to the
  solar value, i.e,
  $[X/H]=\log(N_{X}/N_{H}) - \log(N_{X}/N_{H})_{\odot}$.

  Finally, in Table~\ref{kinematic_table} radial velocities and
  Galactic spatial-velocity components for the
  observed stars with respect to the LSR are presented. 
  The assumed solar motion with respect to the LSR is
  ($U_{\odot}$, $V_{\odot}$, $W_{\odot}$) = (10.0, 5.25, 7.17) km s$^{\rm -1}$ \citep{1998MNRAS.298..387D}.


\onllongtabL{2}{
\begin{center}
\label{parameters_table_full}
\begin{tiny}
\begin{longtable}{llccccccccc}
\multicolumn{11}{c}{{ \tablename\ \thetable{}  Basic physical parameters and metallicities for the stars measured in this work.}}\\
\hline
HIP/          &  HD &  T$_{\rm eff}$ &  $\log g$      & $\xi_{t}$     & [Fe/H] &  $\left\langle A({\rm Fe}~{\rm I}) \right\rangle$ & n$_{\rm I}$ &
 $\left\langle A({\rm Fe}~{\rm II}) \right\rangle$  & n$_{\rm II}$ &  Spec.$^{\dag}$ \\
Other name    &     & (K) & (${\rm cm s^{-2}}$)  & (${\rm km s^{-1}}$) & dex  &      &      &      &      &      \\
 (1)& (2) & (3) & (4)   & (5)            & (6)           & (7)  & (8)  & (9)  & (10) & (11)  \\
\hline
\endfirsthead
\multicolumn{11}{c}{{ \tablename\ \thetable{}  Continued}}\\
\hline
HIP/          &  HD &  T$_{\rm eff}$ &  $\log g$      & $\xi_{t}$     & [Fe/H] &  $\left\langle A({\rm Fe}~{\rm I}) \right\rangle$ & n$_{\rm I}$ &
 $\left\langle A({\rm Fe}~{\rm II}) \right\rangle$  & n$_{\rm II}$ &  Spec.$^{\dag}$ \\
Other name    &     &      (K)   & (${\rm cm s^{-2}}$)  & (${\rm km s^{-1}}$) & dex  &      &      &      &      &      \\
 (1)& (2) & (3) & (4)   & (5)            & (6)           & (7)  & (8)  & (9)  & (10) & (11)   \\
\hline
\endhead
\hline
\endfoot
\endlastfoot
\multicolumn{11}{c}{\textbf{Giants with planets}}\\
\noalign{\smallskip}\hline\noalign{\smallskip}
1692	&	1690	&	4343	$\pm$	20	&	2.06	$\pm$	0.08	&	1.56	$\pm$	0.14	&	-0.23	$\pm$	0.04	&	7.27	$\pm$	0.05	&	197	&	7.27	$\pm$	0.07	&	17	&	2	\\
4297	&	5319	&	4900	$\pm$	25	&	3.35	$\pm$	0.09	&	1.10	$\pm$	0.10	&	0.05	$\pm$	0.04	&	7.55	$\pm$	0.04	&	234	&	7.55	$\pm$	0.06	&	18	&	2	\\
10085	&	13189	&	4175	$\pm$	33	&	1.62	$\pm$	0.13	&	1.49	$\pm$	0.17	&	-0.37	$\pm$	0.06	&	7.13	$\pm$	0.07	&	229	&	7.13	$\pm$	0.10	&	20	&	1	\\
12247	&	16400	&	4864	$\pm$	25	&	2.65	$\pm$	0.08	&	1.42	$\pm$	0.10	&	-0.03	$\pm$	0.03	&	7.47	$\pm$	0.04	&	217	&	7.47	$\pm$	0.06	&	17	&	2	\\
	&	17092	&	4634	$\pm$	28	&	2.48	$\pm$	0.10	&	1.31	$\pm$	0.13	&	0.11	$\pm$	0.05	&	7.61	$\pm$	0.05	&	237	&	7.61	$\pm$	0.08	&	21	&	1	\\
20889	&	28305	&	4915	$\pm$	25	&	2.75	$\pm$	0.08	&	1.40	$\pm$	0.13	&	0.15	$\pm$	0.04	&	7.65	$\pm$	0.05	&	234	&	7.65	$\pm$	0.06	&	23	&	1	\\
36616	&	59686	&	4666	$\pm$	20	&	2.62	$\pm$	0.07	&	1.31	$\pm$	0.11	&	0.13	$\pm$	0.04	&	7.63	$\pm$	0.05	&	234	&	7.63	$\pm$	0.06	&	21	&	1	\\
\hline
\end{longtable}
\end{tiny}
\end{center}
}


\onllongtabL{3}{
\begin{center}
\label{evolutionary_table_full}
\begin{longtable}{lcccccccc}
\multicolumn{8}{c}{{ \tablename\ \thetable{}  Photometric and evolutionary parameters.}}\\
\hline
HIP/       & A$_{\rm V}$ &  $L_{\star}/L_{\odot}$ & T$_{\rm eff}^{\rm phot}$ &  $\log g_{\rm evol}$ &   Age    & Mass           & Radius        \\
Other name & (mag)       &  ($\log$)              &        (K)               &  (${\rm cm s^{-2}}$) &   (Gyr)  & (M$_{\odot}$)  & (R$_{\odot}$)  \\          
 (1)       & (2)         &  (3)                   & (4)                      & (5)                  &   (6)    & (7)            & (8)            \\
\hline
\endfirsthead
\multicolumn{8}{c}{{ \tablename\ \thetable{}  Continued}}\\
\hline
HIP/          & A$_{\rm V}$ &  $L_{\star}/L_{\odot}$ & T$_{\rm eff}^{\rm phot}$ & $\log g_{\rm evol}$ &   Age    & Mass           & Radius          \\
Other name    & (mag)       &  ($\log$)              &        (K)               &  (${\rm cm s^{-2}}$) &   (Gyr)  & (M$_{\odot}$)  & (R$_{\odot}$)   \\    
 (1)          & (2)         &  (3)                   & (4)                      & (5)                  &   (6)    & (7)            & (8)             \\
\hline
\endhead
\hline
\endfoot
\endlastfoot
\multicolumn{8}{c}{\textbf{Giants with planets}}\\
\noalign{\smallskip}\hline\noalign{\smallskip}
1692	&	0.10	$\pm$	0.04	&	1.53	$\pm$	0.39	&    -				&	1.90	$\pm$	0.10	&	6.72	$\pm$	3.18	&	1.11	$\pm$	0.15	&	18.80	$\pm$	2.77	\\
4297	&	0.10	$\pm$	0.03	&	0.96	$\pm$	0.09	&	4960	$\pm$	86	&	3.40	$\pm$	0.07	&	3.45	$\pm$	0.66	&	1.37	$\pm$	0.08	&	3.72	$\pm$	0.36	\\
10085	&	0.82	$\pm$	0.50	&	3.04	$\pm$	0.41	&	4334	$\pm$	168	&	1.40	$\pm$	0.11	&	4.56	$\pm$	2.97	&	1.19	$\pm$	0.25	&	34.60	$\pm$	6.28	\\
12247	&	0.08	$\pm$	0.06	&	1.73	$\pm$	0.04	&	4839	$\pm$	86	&	2.71	$\pm$	0.05	&	1.38	$\pm$	0.18	&	1.90	$\pm$	0.12	&	9.67	$\pm$	0.40	\\
HD 17092	&	0.20	$\pm$	0.05	&	1.15	$\pm$	0.44	&	4504	$\pm$	86	&	2.89	$\pm$	0.28	&	5.60	$\pm$	3.17	&	1.20	$\pm$	0.20	&	6.34	$\pm$	2.18	\\
20889	&	0.06	$\pm$	0.03	&	1.93	$\pm$	0.02	&	4897	$\pm$	80	&	2.66	$\pm$	0.02	&	0.50	$\pm$	0.06	&	2.76	$\pm$	0.08	&	12.39	$\pm$	0.28	\\
36616	&	0.00	$\pm$	0.00	&	1.87	$\pm$	0.02	&	4621	$\pm$	78	&	2.51	$\pm$	0.04	&	1.30	$\pm$	0.32	&	2.02	$\pm$	0.16	&	12.55	$\pm$	0.40	\\
\end{longtable}
\end{center}
}


\onllongtabL{5}{
\begin{center}
\label{tabla_abundancias_full}
\begin{scriptsize}
\begin{longtable}{lrrrrrrrrrrrrrrr}
\multicolumn{16}{c}{{ \tablename\ \thetable{}  Derived abundances of Na, Mg, Al, Si, Ca, Sc, Ti~{\sc i}, Ti~{\sc ii}, V, Cr~{\sc i}, Cr~{\sc ii}, Mn, Co, Ni, and Zn.}}\\
\hline
HIP/Other & [Na/H]& [Mg/H] &  [Al/H] & [Si/H] & [Ca/H] & [Sc/H] & [Ti~{\sc i}/H] & [Ti~{\sc ii}/H] & [V/H] &  [Cr~{\sc i}/H] &  [Cr~{\sc ii}/H] & [Mn/H] & [Co/H] & [Ni/H] & [Zn/H] \\
\hline
\endfirsthead
\multicolumn{16}{c}{{ \tablename\ \thetable{}  Continued}}\\
\hline
HIP/Other & [Na/H]& [Mg/H] &  [Al/H] & [Si/H] & [Ca/H] & [Sc/H] & [Ti~{\sc i}/H] & [Ti~{\sc ii}/H] & [V/H] &  [Cr~{\sc i}/H] &  [Cr~{\sc ii}/H] & [Mn/H] & [Co/H] & [Ni/H] & [Zn/H] \\
\hline
\endhead
\hline
\endfoot
\endlastfoot
\multicolumn{16}{c}{\textbf{Giants with planets}}\\
\noalign{\smallskip}\hline\noalign{\smallskip}
1692         &       -0.12 &       -0.17 &        0.10 &        0.07 &       -0.34 &       -0.11 &        0.12 &        0.12 &        0.23 &       -0.30 &       -0.20 &        0.05 &       -0.20 &       -0.12 &       -0.22\\
 & $\pm$        0.11 & $\pm$        0.06 & $\pm$        0.04 & $\pm$        0.12 & $\pm$        0.12 & $\pm$        0.27 & $\pm$        0.15 & $\pm$        0.22 & $\pm$        0.24 & $\pm$        0.10 & $\pm$        0.13 & $\pm$        0.18 & $\pm$        0.17 & $\pm$        0.09 & $\pm$        0.29\\
4297         &        0.13 &        0.05 &        0.24 &        0.15 &       -0.01 &       -0.05 &        0.15 &        0.09 &        0.32 &       -0.02 &        0.01 &        0.33 &        0.02 &        0.12 &        0.12\\
 & $\pm$        0.06 & $\pm$        0.06 & $\pm$        0.03 & $\pm$        0.06 & $\pm$        0.10 & $\pm$        0.13 & $\pm$        0.04 & $\pm$        0.09 & $\pm$        0.15 & $\pm$        0.04 & $\pm$        0.08 & $\pm$        0.14 & $\pm$        0.11 & $\pm$        0.04 & $\pm$        0.20\\
10085        &       -0.11 &       -0.21 &       -0.07 &       -0.01 &       -0.36 &       -0.09 &        0.15 &       -0.17 &        0.35 &       -0.20 &       -0.40 &       -0.05 &       -0.29 &       -0.19 &       -0.56\\
 & $\pm$        0.09 & $\pm$        0.10 & $\pm$        0.12 & $\pm$        0.09 & $\pm$        0.12 & $\pm$        0.27 & $\pm$        0.16 & $\pm$        0.16 & $\pm$        0.24 & $\pm$        0.12 & $\pm$        0.14 & $\pm$        0.23 & $\pm$        0.16 & $\pm$        0.09 & $\pm$        0.21\\
12247        &        0.16 &        0.05 &        0.11 &        0.13 &       -0.04 &       -0.23 &        0.06 &        0.04 &        0.08 &       -0.08 &       -0.15 &        0.22 &       -0.16 &       -0.01 &        0.02\\
\hline
\end{longtable}
\end{scriptsize}
\end{center}
}


\onllongtabL{7}{
\begin{center}
\label{kinematic_table_full}
\begin{longtable}{lrrrrc}
\multicolumn{6}{c}{{ \tablename\ \thetable{}  Radial velocities and Galactic spatial-velocity components.}}\\
\hline
HIP/          & V$_{\rm r}$$^{\star}$   &  $U_{\rm LSR}$         & $V_{\rm LSR}$            & $W_{\rm LSR}$        &  C$^{\dag}$ \\
Other         & (kms$^{-1}$)            &  (kms$^{-1}$)          & (kms$^{-1}$)             & (kms$^{-1}$)         &                 \\
 (1)          & (2)                     &  (3)                   & (4)                      & (5)                  &   (6)           \\
\hline
\endfirsthead
\multicolumn{6}{c}{{ \tablename\ \thetable{}  Continued}}\\
\hline
HIP/          & V$_{\rm r}$$^{\star}$   &  $U_{\rm LSR}$         & $V_{\rm LSR}$            & $W_{\rm LSR}$        &  C$^{\dag}$ \\
Other         & (kms$^{-1}$)            &  (kms$^{-1}$)          & (kms$^{-1}$)             & (kms$^{-1}$)         &                 \\
 (1)          & (2)                     &  (3)                   & (4)                      & (5)                  &   (6)           \\
\hline
\endhead
\hline
\endfoot
\endlastfoot
\hline\noalign{\smallskip}
\multicolumn{6}{c}{{\bf Giants with planets}}\\
\hline\noalign{\smallskip}
1692	&	17.58	$\pm$	0.25	&	-8.80	$\pm$	8.15	&	3.94	$\pm$	3.75	&	-10.75	$\pm$	0.94	&	D	\\
4297	&	-0.21	$\pm$	0.31	&	25.28	$\pm$	1.68	&	-13.56	$\pm$	2.00	&	-5.40	$\pm$	1.33	&	D	\\
10085	&	25.36	$\pm$	0.46	&	-13.81	$\pm$	3.55	&	22.85	$\pm$	3.10	&	8.97	$\pm$	6.33	&	D	\\
12247	&	8.60	$\pm$	0.31	&	3.65	$\pm$	0.49	&	-20.57	$\pm$	1.33	&	-0.63	$\pm$	0.41	&	D	\\
HD17092	&	5.56	$\pm$	0.31	&	-7.50	$\pm$	6.73	&	-7.26	$\pm$	8.17	&	8.42	$\pm$	1.49	&	D	\\
20889	&	38.21	$\pm$	0.23	&	-31.04	$\pm$	0.23	&	-13.42	$\pm$	0.27	&	6.02	$\pm$	0.19	&	D	\\
36616	&	-36.17	$\pm$	0.35	&	57.99	$\pm$	0.56	&	-18.66	$\pm$	1.07	&	0.29	$\pm$	0.32	&	D	\\
\end{longtable}
\end{center}
}


\end{document}